\documentclass[a4paper,11pt]{article}
\pdfoutput=1 

\usepackage{jcappub} 

\usepackage{soul}

\newcommand{\T}{\mathbf{\hat{\mathcal{T}}}}

\newcommand{\Dk}[1]{\frac{d^3#1}{(2\pi)^3}}

\newcommand{\ve}[1]{{\text{\bf #1}}} 
\newcommand{\vk}{\ve k}
\newcommand{\vp}{\ve p}
\newcommand{\vq}{\ve q}
\newcommand{\vx}{\ve x}

\newcommand{\vr}{\ve r}

\newcommand{\mA}{\mathcal{A}}
\newcommand{\mB}{\mathcal{B}}

\newcommand{\A}{\mathcal{A}}
\newcommand{\B}{\mathcal{B}}

\newcommand{\dD}{\delta_\text{D}}

\title{\boldmath 
Marked correlation functions in perturbation theory}

\author[a,b]{Alejandro Aviles,}
\emailAdd{avilescervantes@gmail.com}

\author[c]{Kazuya Koyama,}

\author[b]{Jorge L. Cervantes-Cota,}

\author[d,c]{Hans A. Winther,}

\author[e]{Baojiu Li}

\affiliation[a]{Consejo Nacional de Ciencia y Tecnolog\'ia, Av. Insurgentes Sur 1582,
Colonia Cr\'edito Constructor, Del. Benito Jurez, 03940, Ciudad de M\'exico, M\'exico}
\affiliation[b]{Departamento de F\'isica, Instituto Nacional de Investigaciones Nucleares,
Apartado Postal 18-1027, Col. Escand\'on, Ciudad de M\'exico, 11801, M\'exico.}

\affiliation[c]{
Institute of Cosmology and Gravitation, University of Portsmouth, Portsmouth, PO1 3FX, UK}

\affiliation[d]{Institute of Theoretical Astrophysics, University of Oslo, 0315 Oslo, Norway}

\affiliation[e]{Institute for Computational Cosmology,  Department of Physics, Durham University, South Road, Durham DH1 3LE, UK}


\abstract{
We develop perturbation theory approaches to model the marked correlation function constructed to up-weight low density regions of the Universe, which might help distinguish modified gravity models from the standard cosmology model based on general relativity. Working within Convolution Lagrangian Perturbation Theory, we obtain that weighted correlation functions are expressible as double convolution integrals that we approximate using a combination of Eulerian and Lagrangian schemes. We find that different approaches agree within 1$\%$ on quasi non-linear scales. Compared with {\it N}-body simulations, the perturbation theory is found to provide accurate predictions for the marked correlation function of dark matter fields, dark matter halos as well as Halo Occupation Distribution galaxies down to $30$ Mpc/h. These analytic approaches help to understand the degeneracy between the mark and the galaxy bias and find a way to maximize the differences among various cosmological models.
}

\begin{document}

 \maketitle

\flushbottom

 

\begin{section}{Introduction}

In studying the clustering of objects in the sky, the most important statistics is the two-point correlation function, or its counterpart in Fourier space, the power spectrum. The reason of this is the nearly Gaussian nature of linear fluctuations in the early stages of the
Universe's evolution. This method of neglecting nonlinear evolution has proved to be extremely successful for explaining the anisotropies of the cosmic microwave background radiation \cite{Aghanim:2018eyx}. However, nonlinearities are inherent to gravitational instability, and they dominate the evolution of small scales at low redshift, such that the incorporation of higher-than-leading order contributions to matter fluctuations has been 
critical for the analytical understanding, and the construction of theoretical templates, of the processes that yield the structures we observe nowadays \cite{Bernardeau:2001qr}. Despite the success of perturbation theory (PT), the improvement over linear theory is only important at the edge of the linear regime, named the mildly nonlinear regime. Soon after nonlinear evolution is completely onset, field fluctuations become badly approximated by PT and their predictions become non-trustable. At this point, the direct, brute force and computationally expensive approach of {\it N}-body simulations is the most useful tool to study the clustering of 
dark matter and tracers, leading to the ``true'' solution of the problem. Another promising route to study the highly non-Gaussian processes is given by effective field theory, where small scales physics are integrated out of the theory and their impact over the large scale incorporated under a set of free parameters to be determined by observations \cite{Baumann:2010tm}.

A different, complementary route to the study of large-scale structure formation, which is the subject of study in this work, relies on statistics that by construction consider mainly linear fields, such that nonlinearities become subdominant at any time of the matter clustering. To this end we construct weighted correlation functions that up-weights low density regions in the Universe \cite{White:2016yhs}. The process to do so is to define an algebraic function -- the mark function -- that assigns a mark to each object under consideration (it could be any tracer of the underlying matter density, or even the dark matter field itself), which thereafter weights the point process on which we are interested to compute statistics; for definiteness, the two-point statistics in configuration space. 
Thereafter, a marked correlation function 
\cite{Beisbart:2000ja,Beisbart_2002,Gottloeber:2002vm,Sheth:2005ai,Sheth:2005aj,Skibba:2005kb,White:2008ii,White:2016yhs} is constructed by factorizing the clustering of ``unmarked'' tracers and the mean mark obtained by averaging the mark of each of the sampled objects. In this way one isolates the effects of the mark, and focuses only on the clustering of marks.  

Marked statistics have by now a long history \cite{Beisbart:2000ja}; they have been used to assign properties to objects, such as the luminosity, color, and morphology of galaxies \cite{Sheth:2005ai,Sheth:2005aj}, and to break degeneracies between Halo Occupation Distribution (HOD) and cosmologies that arise because one is 
usually able to redistribute the galaxies by compensating the halo mass function in order to obtain the same two-point correlation function \cite{White:2008ii}.

On the other hand, since the discovery of the accelerated expansion of the Universe \cite{Perlmutter:1998np,Riess:2004nr}, tons of models that modify Einstein gravitational theory in the infrared, generically called modified gravity (MG), were constructed in order to 
explain the speeding up in the background expansion rate, 
as an alternative to the 
$\Lambda$CDM model, see refs.~\cite{Horndeski:1974wa,Dvali:2000hr,Capozziello:2002rd,Carroll:2003wy,Nicolis:2008in} among 
many others. The main difficulty of constructing such models is, of course, to maintain the success of general relativity (GR). In particular, in order to 
not spoil Newtonian gravity in describing a wide variety of astronomical observations, MG models rely on nonlinear mechanisms that effectively screen their associated fifth-forces in high density or 
strong gravitational potential regions \cite{Vainshtein:1972sx,Khoury:2003aq,Khoury:2003rn}. In such a way, MG can have an important impact on the cosmological 
scales and low density regions, while 
complying with observations at higher densities; see e.g., refs.~\cite{Clifton:2011jh,Koyama:2015vza,doi:10.1142/11090} for recent reviews. Motivated by these nonlinear screening effects, 
Ref.~\cite{White:2016yhs} proposed the use of marked correlation functions that up-weight low-density, close-to-linear regions of space; which were further studied in refs.~\cite{Valogiannis:2017yxm,Hernandez-Aguayo:2018yrp,Armijo:2018urs,Satpathy:2019nvo} with MG numerical simulations. The main and first step is to mark the observed objects with a function, the ``mark function'' $m$, that smoothly under-represents them as they reside in regions with higher and higher densities. In this way, the objects with larger assigned marks (or larger weights) are enhanced in statistics, where the screening mechanism becomes less important, and the MG effects, hopefully, may be captured more neatly.

In this article, building upon the work of \cite{White:2016yhs}, we will use first linear standard perturbation theory, but given that the ``standard" correlation function is better modeled by Lagrangian perturbation theory (LPT), we will develop the marked correlation function theory in this frame, more precisely with Convolution-LPT (CLPT) \cite{Carlson:2012bu}, as well. Contrary to standard perturbation theory in MG \cite{2009PhRvD..79l3512K}, the MG LPT for matter and tracers in large scale structure formation has been developed until recently 
\cite{Aviles:2017aor,Winther:2017jof,Aviles:2018saf,Aviles:2018qot,Valogiannis:2019xed,Valogiannis:2019nfz} and applied to 
hybrid {\it N}-body/analytical treatments in refs.~\cite{Valogiannis:2016ane,Winther:2017jof,Moretti:2019bob}. However, the methods developed on the different schemes can be translated by means of a set of kernel identifications between Lagrangian and Eulerian frames \cite{Aviles:2018saf}.

During the development of the CLPT marked correlation functions theory we will clarify some points in the work of ref.~\cite{White:2016yhs}. Namely, the relation of the mark
Taylor coefficients, $C$, with the 
resummed expansion parameters $B$. Furthermore, the exact CLPT of weighted correlation functions leads to a double three-dimensional integral convolution of matter fluctuations and biased tracers [see eq.~(\ref{mCFCLPT2}) in section~\ref{subsec:mCFCLPT}], that is reduced to a single convolution in \cite{White:2016yhs} by identifying Eulerian coordinates of linear fields with Lagrangian coordinates. This is a consequence, and a drawback, of the use of LPT, because it relies on evolving initial yet linear matter and tracers densities, but, on the other hand,  the marks are assigned at the moment of observation with the use of already evolved Eulerian densities. And hence, either the use of mixing Eulerian-Lagrangian methods, or the reduction to pure Lagrangian methods as in ref.~\cite{White:2016yhs}, is unavoidable.  Therefore, we also propose a method by which some ingredients are kept exact within CLPT, while others use of resummations that bring them in a form close to SPT. When comparing all the used methods, in this work, we find that they differ by less than the 1\%, even for methods that do not include loop corrections. This shows that the use of linear theory, which is simpler for modelling and coding, is reliable for some applications of marked statistics.

We will exemplify our findings with the representative Hu-Sawicki $f(R)$ gravity model \cite{Hu:2007nk}; specifically we will use the F6, F5 and F4 models, that are introduced in appendix \ref{app:HS}. Our findings for GR make use of the exact kernels in $\Lambda$CDM, and not the widely used static-approximation, EdS kernels. For our analytical findings we make use of a modified version of the publicly available code 
\texttt{MGPT},\footnote{\href{https://github.com/cosmoinin/MGPT}{https://github.com/cosmoinin/MGPT}} that computes kernels and integrates a set of functions necessary to build the correlation functions and power spectra for biased tracers at 1-loop in MG; see ref.~\cite{Aviles:2018saf} for details. All our results are shown at redshift $z=0.5$ and we use WMAP 9yr best fit parameters given by $\Omega_b = 0.046$, $\Omega_m = 0.281$, $h = 0.697$, $\sigma_8 = 0.82$ and $n_s = 0.971$ \cite{Hinshaw:2012aka}.


The rest of this work is organized as follows: In section~\ref{sec:Up} we define the mark correlation function and tracers of the dark matter field, in section~\ref{sec:Eulerian} we formally introduce the mark, explain the renormalized bias and establish the relationship of the mark Taylor expansion parameters ($C$'s) and the resummed expansion  parameters ($B$'s) to be able to compute the Euleran correlation of weighted tracers and its corresponding marked correlation function.  In  section~\ref{sec:Lag_space} we compute the marked correlation function in CLPT, and in order to integrate it we use an approximation in section~\ref{subsec:mCFCLPTapprox}.  In section~\ref{sect:Degeneracies} we discuss the effects and degeneracies of bias and mark parameters. In section \ref{sec:CS} we compare our analytical results to $\Lambda$CDM and Hu-Sawicki $f(R)$ 
models making use of Extended LEnsing PHysics using ANalaytic ray Tracing ({\sc elephant}) simulations \citep{Cautun:2017tkc}. In section~\ref{sec:Conclu} we further discuss and conclude on our perturbative approach to marks. Finally, in the appendices we added complementary material on the general theory of LPT for MG (appendix \ref{app:LPT}), cumbersome formulae of the CLPT formalism to 1-loop and approximations (appendix \ref{app:mCF}), the computation of the different statistics needed in the main text formulae (appendix \ref{app:kqfunctions}), and finally we put forward a way to add curvature and tidal bias to the formalism (appendix \ref{subsec:curvbias}).   

\end{section}

 \begin{section}{Up-weighting low density regions with marked correlation functions}\label{sec:Up}
 
A marked correlation function (mCF) \cite{Beisbart:2000ja,Gottloeber:2002vm,Sheth:2005ai,Sheth:2005aj,Skibba:2005kb,White:2008ii,White:2016yhs} is defined as the sum of pairs of objects separated by a distance $r$, weighted by the ratio of the mark function value to the mean mark $m_i/\bar{m}$ at each point and  divided by the number of pairs $n(r)$:
\begin{equation}\label{mCF}
 \mathcal{M}(r) = \sum_{ij|r_{ij}=r}\frac{m_i m_j}{n(r)\bar{m}^2}.
\end{equation}
That is, it is a 2-point statistics of the clustering of marks. 
We choose the mark to be a function of the environmental matter density in which such objects reside, smoothed over a scale that we take to be larger than the size of the objects. We should note that in real applications, one rarely has the dark matter densities at hand, so we will relax this assumption later so that our formalism can be applied to halos and galaxy mocks.   
In viable chameleon modified gravity 
theories, high energy density regions are screened and these models reduce to General Relativity. 
Hence, the effects of MG are expected to be more pronounced in low density regions. For this reason marked functions that enhance low
density regions have been considered in the MG literature. We write such a relation as 
\begin{equation}
m(\rho)  = G[ \delta_R(\vx)], 
\end{equation}
with $\delta_R(\vx,t) = \int d\vx' W_R(|\vx-\vx'|/R) \delta(\vx') $ 
a smoothed matter overdensity and $R$ the smoothing scale.  The mark  function (hereafter, the White-mark) 
\begin{equation} \label{MWmark}
m= \left(\frac{ 1+ \rho_*}{1+ \rho_* + \delta_R} \right)^p
\end{equation}
is proposed in ref.~\cite{White:2016yhs}, with $p$ and $\rho_*$ dimensionless parameters chosen in order to up-weight low density regions. 
Other marks have been also proposed; see refs.~\cite{Valogiannis:2017yxm,Hernandez-Aguayo:2018yrp,Armijo:2018urs}.

The objects we consider are tracers of the dark matter field $\delta$. Matter and tracers $X$ overdensities are
related through the bias function\footnote{This relation has a stochastic nature, 
that can be made explicit by writing $F_\vx[\delta(\vx);\vx]$ \cite{Schmidt:2012ys}. Dependencies on other bias operators can be also introduced
as arguments in $F$.} 
\begin{equation} \label{defFx}
 1+\delta_X(\vx) = F_\vx[\delta(\vx)].
\end{equation}
It is known that a well defined bias expansion within chameleon theories must contain higher-order derivative operators ($\nabla^2\delta$, $\nabla^4\delta$, and so on); see e.g. section 8.3 of \cite{Desjacques:2016bnm} and ref.~\cite{Aviles:2018saf}. 
Hence, we will introduce curvature bias, as well as tidal bias, later
in appendix \ref{subsec:curvbias}.
In order for some quantities to be well defined, we also smooth matter overdensities with a scale  $R_\Lambda$ 
that we take to be ideally $R_\Lambda \ll R$, hence we simply denote $\delta = \delta_\Lambda$; at the end, in
numerical calculations we set $R_\Lambda=0$. 

\end{section}

\begin{section}{Perturbative treatment of the marked correlation function in Eulerian space} \label{sec:Eulerian}

For a perturbative treatment we expand the mark in a power series of the smoothed overdensity
\begin{equation}\label{Taylorm}
m(\delta_R; C_i)  =C_0 + C_1 \delta_R + \frac{1}{2} C_2 \delta^2_R + \cdots
\end{equation}
with $C_i =  G^{(i)}[0]$, the $i$th derivative of $G[\delta_R]$ evaluated at $\delta_R=0$. These parameters are intended to 
enhance low density regions where screenings are not very efficient and the effects of MG might be detected. 
For the White-mark in eq.~(\ref{MWmark}) one obtains $C_0=1$, $C_1 = -p/(1+\rho_*)$, $C_2 = p(p+1)/(1+\rho_*)$. The fewer parameters are needed to model the mark, the better the convergence will be. 
We notice that $C_1<0$  enhances low density regions.

It is well known that the correlation function is better modeled by LPT (see e.g. ref.~\cite{Baldauf:2015xfa}), 
however to gain insight and as a  warm-up for Lagrangian 
calculations, in this section we compute the marked correlation function in Eulerian space.
The mean mark is given by mark weighted by the tracers density field, 
\begin{align} \label{prebarmE}
\bar{m} &= \langle G[\delta_R(\vx)](1+ \delta_X(\vx)) \rangle \nonumber\\
&= \Big \langle  \Big(C_0+C_1 \delta_R(\vx) + \frac{C_2}{2}\delta^2_R(\vx) + \cdots \Big)
\Big(c_0+c_1 \delta(\vx) + \frac{c_2}{2}\delta^2(\vx) + \cdots \Big) \Big \rangle \nonumber\\
&=c_0C_0+ c_1C_1 \sigma^2_R  + \frac{1}{2}c_0C_2 \sigma^2_{RR}  + \frac{1}{2}C_0c_2 \sigma^2 + \cdots,
\end{align}
with $c_i = F_\vx^{(i)}[0]$, and zero-lag correlators defined as
\begin{equation}
\sigma^2 \equiv \langle (\delta(0))^2\rangle, \quad \sigma^2_R \equiv \langle \delta_R(0)  \delta(0)\rangle, \quad \sigma^2_{RR} \equiv \langle(\delta_{R}(0))^2\rangle.
\end{equation}
Below, we use also the correlation and cross-correlation functions
\begin{equation}
\xi(r) \equiv \langle \delta(\vx)\delta(\vx+\vr)\rangle, \quad \xi_R(r) \equiv \langle \delta_R(\vx)\delta(\vx+\vr)\rangle, \quad \xi_{RR}(r) 
\equiv \langle\delta_R(\vx)\delta_R(\vx+\vr)\rangle.
\end{equation}
Now, it is convenient to write eq.~(\ref{prebarmE}) in terms of renormalized bias parameters \cite{Matsubara:2008wx}
\begin{equation}\label{bnEdef}
 b_n^E = \int \frac{d\lambda}{2\pi} e^{-\lambda^2 \sigma^2/2} \tilde{F_\vx}(\lambda) (i\lambda)^n, 
\end{equation}
where $\tilde{F}_\vx(\lambda)$ is the Fourier transform of $F_\vx$ with spectral parameter $\lambda$ (dual to $\delta$). 
This definition of bias is more commonly used in LPT, so we are using  
the label ``$E$'' to distinguish Eulerian from Lagrangian biases. Equation (\ref{bnEdef}) leads to the 
precise resummation of bare bias parameters $c_n$ that yields the renormalized bias parameters, becoming related by $b_n^E=\sum_{k=0}^{\infty} c_{n+2k} \sigma^{2k}/(2^k k!)$ \cite{Aviles:2018thp,Eggemeier:2018qae}. 
Analogously, we introduce the ``resummed'' expansion parameters $B_n$ as
\begin{equation}\label{Bndef}
B_n=\frac{B_n^*}{B_0^*} \qquad \text{with} \qquad B_n^* = \int \frac{d\Lambda}{2\pi} e^{-\Lambda^2 \sigma^2_{RR}/2} \tilde{G}(\Lambda) (i\Lambda)^n,
\end{equation}
where $\tilde{G}(\Lambda)$ is the Fourier transform of $G$, and $\Lambda$ is a spectral parameter, dual to $\delta_R$,
and find that $B_n^*$ and $C_n$ are related by
$B_n^*=\sum_{k=0}^{\infty} C_{n+2k} \sigma^{2k}_{RR}/(2^k k!)$. Therefore, the expansion parameters and the Taylor coefficients of the mark function relate as
\begin{equation}\label{BnRenorm}
B_n(C_n,\sigma_{RR}^2) = \frac{\sum_{k=0}^{\infty} C_{n+2k} \sigma^{2k}_{RR}/(2^k k!)}{\sum_{k=0}^{\infty}  C_{2k} \sigma^{2k}_{RR}/(2^k k!)}.
\end{equation}
Inserting the $b_n^E$ and $B_n$ parameters in eq.~(\ref{prebarmE}) we get
\begin{equation} \label{barmE}
 \bar{m} = B_0^* \big[ 1 + b_1^E B_1 \sigma^2_R + \cdots \big],
\end{equation}
which is the mean mark reported in ref.~\cite{White:2016yhs}. We have defined the expansion parameters in this way because the factor $B_0^*$ is absorbed by the mean mark and we do not have to carry it in all expressions for the marked correlation function. Moreover, typically $C_0=1$, and one smooths over large regions, such that $\sigma^2_{RR} \ll 1$ and $B_n \simeq B_n^* \simeq C_n$.

We compute the correlation of weighted (by the mark function $m$) tracers separated by a distance $r=|\vx_2-\vx_1|$, 
\begin{align}
&\langle G_{1} F_{\vx,1} G_{2} F_{\vx,2}\rangle  \equiv \langle G[\delta_R(\vx_1)]\big(1+\delta_X(\vx_1)\big)   G[\delta_R(\vx_2)]\big(1+\delta_X(\vx_2)\big) \rangle 
\nonumber \\
&= \int \frac{d\lambda_1 d\lambda_2 d\Lambda_1  d\Lambda_2}{(2 \pi)^4} \langle e^{ i(\lambda_1\delta_1 +\Lambda_1\delta_{R,1} + \lambda_2\delta_2 +\Lambda_2\delta_{R,2})} \rangle \tilde{F}_\vx(\lambda_1)\tilde{F}_\vx(\lambda_2)
\tilde{G}(\Lambda_1)\tilde{G}(\Lambda_2) \nonumber\\
&= \int \frac{d\lambda_1 d\lambda_2 d\Lambda_1  d\Lambda_2}{(2 \pi)^4} \tilde{F}_\vx(\lambda_1)\tilde{F}_\vx(\lambda_2)
\tilde{G}(\Lambda_1)\tilde{G}(\Lambda_2) 
 e^{ -\frac{1}{2}(\lambda_1^2 + \lambda_2^2)\sigma^2 -\frac{1}{2}(\Lambda_1^2 + \Lambda_2^2)\sigma^2_{RR} } \nonumber\\
 &\quad \times \Big[1 - (\lambda_1\Lambda_1 + \lambda_2\Lambda_2 )\sigma^2_R   - \lambda_1\lambda_2 \xi(r) -\Lambda_1\Lambda_2 \xi_{RR}(r) - 
 (\lambda_1\Lambda_2+\lambda_2\Lambda_1)\xi_R(r) + \cdots \Big] \nonumber\\
&= \bar{m}^2\Big[ 1+(b_1^E)^2 \xi(r)+B_1^2 \xi_{RR}(r) + 2 b_1^E B_1 \xi_{R}(r) +\cdots \Big],
\end{align}
where the ellipsis denote second order terms in $\xi_{{},R,RR}(r)$. In the above equation $G_{i}$ and $F_{\vx,i}$ denote 
$G[\delta_{R}(\vx_i)]$ and $F_{\vx}[\delta(\vx_i)]$ and we
have shifted to Fourier space ($\delta \rightarrow \lambda$, $\delta_R \rightarrow \Lambda$) in the second equality. In the third equality we used
the cumulant expansion theorem, 
\begin{equation} \label{CET}
 \langle e^{iX} \rangle = \exp\left[ \sum_{N=1}^\infty \frac{i^N}{N!} \langle X^N \rangle_c \right],
\end{equation}
and expanded out of the exponential all terms but those containing
$\sigma^2$ and $\sigma^2_{RR}$, such that we can use eqs.~(\ref{bnEdef}) and (\ref{Bndef}) to get biases from spectral 
parameters, as we did in the last equality.
The mCF is obtained by 
multiplying the above equation by $\bar{m}^{-2}$ and by taking the ratio to the correlation function of tracers. We obtain
\begin{equation} \label{mCFE}
\mathcal{M}^E(r) =  \frac{1+(b_1^E)^2 \xi(r)+B_1^2 \xi_{RR}(r) + 2 b_1^E B_1 \xi_{R}(r) + \cdots}{1+(b_1^E)^2 \xi(r) + \cdots} \equiv \frac{1+W(r)}{1+\xi_X(r)},
\end{equation}
such that the weighted correlation function is given by
\begin{equation}\label{WCF}
1+W(r) = \frac{1}{\bar{m}^2} \langle G_{1} F_{\vx,1} G_{2} F_{\vx,2}\rangle.
\end{equation}
We emphasize that we are weighting the tracer overdensities $F_\vx$ with the mark computed using the total matter 
smoothed overdensities.
We  notice that the zero-lag correlators $\sigma^2$ and $\sigma^2_{RR}$ do not appear in 
the mCF as is guaranteed because we are using renormalized $b$ and $B$ parameters. 
Meanwhile, the cross-covariance $\sigma^2_R$ is canceled out by the mean mark squared appearing 
in the definition of the mCF in eq.~(\ref{mCF}).  We remark that a  process of renormalization of the Taylor expansion
coefficients $C$ is not strictly necessary because the scale $R$ is not arbitrary, but is chosen from the 
beginning to assing the mark to the tracers. However, we have proceeded by doing so, in order to have a simpler structure
of equations that do not carry the variances $\sigma_{RR}^2$ and matches the notation of ref.~\cite{White:2016yhs}.

 As mentioned previously, in real applications, we have to use a mark computed by the number density of galaxies. An advantage of using the renormalized $B$ parameters is that 
the effect of this reassignment can be included simply by re-scaling the $B$ parameters. For the application to dark matter halos and galaxies, we will treat them as free parameters and we will fit these parameters from simulations. 

 \end{section}

 \begin{section}{Perturbative treatment of the marked correlation function in Lagrangian space}\label{sec:Lag_space}

In Lagrangian space one considers regions of space, at an initial early time with spatial coordinates $\vq$, 
that give rise to tracers. We relate matter and tracers overdensities, in an analogous way to eq.~(\ref{defFx}), by
\begin{equation}\label{defF}
 1+\delta_X(\vq) = F[\delta(\vq)].
\end{equation}
The function $F_\vx$, introduced in eq.~(\ref{defFx}), and $F$ are not simply related, 
but assuming number conservation of tracers, $(1+\delta_X(\vx))d^3x = (1+\delta_X(\vq))d^3q$, one obtains
\begin{equation}\label{FxToF}
 F_\vx[\delta(\vx)] = \int \Dk{k} \int d^3 q e^{i\vk\cdot(\vx-\vq)} \int\frac{d\lambda}{2\pi} \tilde{F}(\lambda) e^{i\lambda\delta(\vq) - i \vk\cdot \mathbf{\Psi(\vq,t)}},
\end{equation}
with $\mathbf{\Psi}(\vq)$ the Lagrangian displacement vector, that maps Lagrangian coordinates $\vq$ to Eulerian coordinates $\vx$ as $\vx(\vq,t) = \vq + \mathbf{\Psi}(\vq,t)$.
Equivalently to eq.~(\ref{bnEdef}), we introduce the renormalized Lagrangian local biases \cite{Matsubara:2008wx} with
\begin{equation}\label{bndef}
 b_n = \int \frac{d\lambda}{2\pi} e^{-\lambda^2 \sigma^2/2} \tilde{F}(\lambda) (i\lambda)^n. 
\end{equation}
In this section we evolve initially biased tracers with overdensity $\delta_X(\vq)$ 
using Convolution Lagrangian perturbation theory 
and thereafter assign
them a mark $m[\delta_R(\vx)]$ at the moment of observation, where $R$ is the physical 
size of the (Eulerian) region that hosts the objects and that is chosen from the beginning.
The use of the Lagrangian approach has some advantages 
with respect to the Eulerian. In the first place it is 
well known that the two-point correlation function is poorly 
modeled within the Eulerian approach, particularly at the BAO peak position; 
second, the (renormalized) Lagrangian bias parameters are obtained through 
the peak background split prescription \cite{Kaiser:1984sw,Mo:1996cn,Sheth:1999mn,Schmidt:2012ys}, and hence are physically appealing. But the price to pay is that the mCF is cumbersome to compute, and one must evaluate several 6-dimensional integrals [eq.~(\ref{mCFCLPT2}) below]. However, we will assume that the smoothing scale is large, $R>1/k_\text{NL}$, and that the mark efficiently up-weights regions of low density, such that we can deal with the smoothed matter overdensity $\delta_R$ as a linear field; moreover, in subsection \ref{subsec:mCFCLPTapprox} we propose a model
to approximate the mCF, and we compare our results to ref.~\cite{White:2016yhs}, finding a good agreement between both methods.

\begin{subsection}{The mean mark}\label{subsec:meanmark}

The mean mark can be computed in Lagrangian space as
\begin{align}
  \bar{m} &= \langle G[\delta_R(0)]F_\vx[\delta_X(0)]\rangle =  
  \int \Dk{k} d^3 q \,e^{-i\vk\cdot\vq} \int\frac{d\Lambda}{2\pi} \frac{d\lambda}{2\pi} 
  \tilde{G}(\Lambda)\tilde{F}(\lambda) \langle e^{i\Lambda\delta_R(0) + i\lambda\delta(\vq) - i \vk\cdot \mathbf{\Psi}(\vq,t)} \rangle,
\end{align}
where we used eq.~(\ref{FxToF}) to relate Lagrangian and Eulerian tracer fluctuations. We now
use the cumulant expansion theorem [eq.~(\ref{CET})] with $X=\Lambda\delta_R(0) + \lambda\delta(\vq) -  \vk\cdot \mathbf{\Psi}(\vq,t)$. We have
\begin{align}
-\frac{1}{2}\langle X^2 \rangle_c &= -\frac{1}{2}\Lambda^2 \sigma^2_{RR} - \frac{1}{2}\lambda^2\sigma^2  
- \lambda \Lambda \xi_R(q) + \Lambda k_i U^R_i(\vq)   -\frac{1}{2}k^2 \sigma_\Psi^2 , 
\end{align}
with the dispersion of Lagrangian displacements
\begin{equation}
\sigma_\Psi^2(t) = \frac{1}{3}\delta_{ij}\langle \Psi_i(0) \Psi_j(0)\rangle_c = \frac{1}{6\pi^2} \int_0^\infty dp P_L(p,t), 
\end{equation}
where $P_L(k)$ is the linear matter power spectrum, and 
\begin{equation}
U^R_i(\vq) = \langle \delta_R(0)\Psi_i(\vq) \rangle_c = \int \Dk{p} e^{i\vp\cdot\vq}\frac{p_i}{p^2} \tilde{W}_R(p) P_L(p),
\end{equation}
where the second equality is valid to linear order.  Expanding
correlators at finite separation and using eqs.~(\ref{Bndef}) and (\ref{bndef}) we have
\begin{align} \label{mmLPT}
  \bar{m} &=
 B_0^* \int \Dk{k} \int d^3 q e^{-i\vk\cdot\vq} e^{-k^2\sigma_\Psi^2/2} \Big[ 1 +b_1 B_1 \xi_R(q) -  i B_1  k_i U^R_i(\vq)  +\cdots \Big]\nonumber\\  
  &= B_0^* \Bigg[ 1+ (1+b_1)B_1 \int \Dk{k} e^{-k^2 \sigma_\Psi^2/2}\tilde{W}_R(k) P_L(k) + \cdots \Bigg] \nonumber\\  
  &= B_0^* \Bigg[1+ (1+b_1)B_1 \sigma^2_{R\sigma_\Psi}  + \cdots \Bigg],
\end{align}
where 
we defined
\begin{equation} 
 \xi_{R\sigma_\Psi}(r) \equiv  \int \Dk{k} e^{i\vk\cdot \vr} e^{-k^2 \sigma_\Psi^2/2}\tilde{W}_R(k) P_L(k), \qquad  \sigma^2_{R\sigma_\Psi} \equiv \xi_{R\sigma_\Psi}(0).
\end{equation} 
For large $R$, or at early times, $\sigma_\Psi \ll R$ and we get $\bar{m}= B_0^* \big[ 1 + (1+b_1)B_1 \sigma^2_R \big]$, which is eq.~(\ref{barmE})
with the proper identification $b_1^E = 1+ b_1$. At late times, however, $\sigma_\Psi$ 
grows and becomes important, making the mean mark evolve at a slower rate.
Physically, the result of eq.~(\ref{mmLPT}) follows from the fact that we 
are correlating a linearly evolved smoothed matter density field $\sim\!\tilde{W}_R\delta$ with a Lagrangian evolved field 
$\sim\!e^{-k^2 \sigma^2_\Psi/2}\delta$.\footnote{The Zeldovich approximation matter power spectrum is roughly $P_\text{ZA}(k) \approx e^{-k^2 \sigma^2_\Psi}P_L(k)$ \cite{Matsubara:2007wj}.}

\end{subsection}

\begin{subsection}{The marked correlation function in CLPT}\label{subsec:mCFCLPT}

The weighted correlation function of biased tracers (weighted by the mark function over the mean mark), is given by 
eqs.~(\ref{WCF}). Using eq.~(\ref{FxToF}) to substitute Eulerian by Lagrangian tracers, 
and transforming $G[\delta_R(\vx_{1,2})]$ to Fourier space, we have
\begin{align} \label{mCFCLPT0}
 \langle G_{1}F_{\vx,1}G_{2}F_{\vx,2} \rangle &=
 \int \Dk{k_1}\Dk{k_2} \int d^3q_1 d^3q_2 \int \frac{d\lambda_1}{2\pi}\frac{d\lambda_2}{2\pi}\frac{d\Lambda_1}{2\pi}\frac{d\Lambda_2}{2\pi}
 \tilde{F}(\lambda_1)\tilde{F}(\lambda_2)\tilde{G}(\Lambda_1)\tilde{G}(\Lambda_2) \nonumber\\
 & \qquad e^{i\vk_1\cdot (\vx_1-\vq_1)}e^{i\vk_2\cdot (\vx_2-\vq_2)} \langle e^{i (\lambda_1\delta(\vq_1) + \lambda_2\delta(\vq_2)+\Lambda_1\delta_R(\vx_1)+\Lambda_2\delta_R(\vx_2))}\rangle.
\end{align}
As a standard practice we use the cumulant expansion theorem and expand out of the exponential 
all terms containing spectral parameters, with the exception of 
variances $\sigma^2$ and $\sigma_{RR}^2$, and linear terms in the Lagrangian displacements. 
Performing the $\lambda$ and $\Lambda$ integrations with the aid of eqs.~(\ref{Bndef}) and (\ref{bndef}) we obtain
\begin{align} \label{mCFCLPT1}
\langle G_{1}F_{\vx,1}G_{2}F_{\vx,2} \rangle &=
\int \Dk{k_1}\Dk{k_2} \int d^3q_1 d^3q_2 
 e^{i\vk_1\cdot (\vx_1-\vq_1)}e^{i\vk_2\cdot (\vx_2-\vq_2)} \nonumber\\
 & \qquad e^{-\frac{1}{2}(\vk_1+\vk_2)^2\sigma^2_\Psi} 
 e^{\frac{1}{2}k_1^ik_2^jA^L_{ij}(\vq)} (1 + I)
\end{align}
where the ``1''  gives the Zeldovich Approximation standard  correlation function and the function $I=I(\vx_1,\vx_2, \vq_1,\vq_2,\vk_1,\vk_2)$ contains the biased and marked densities 
and correlations of non-linear Lagrangian displacements. 
A complete expression for the function $I$ up to 1-loop in fluctuations and third order in bias expansion is given in eq.~(\ref{I0}) of appendix \ref{app:mCF}. The matrix \cite{Carlson:2012bu}
\begin{equation}\label{AijL}
 A_{ij}^L(\vq) = \int \Dk{p} \big( 2 - e^{i\vp\cdot \vq} - e^{-i\vp\cdot \vq} \big) \frac{p_ip_j}{p^4}P_L(p)
\end{equation} 
is the leading order piece of the correlation of displacement fields $\langle \Delta_i\Delta_j\rangle_c$, with $\Delta_i = \Psi_i(\vq_2)-\Psi_i(\vq_1)$.
We now redefine variables 
 \begin{align}
  \vq =\vq_2 - \vq_1, &\quad \ve Q = \frac{1}{2}(\vq_1+\vq_2),  \nonumber\\
  \vr =\vx_2 - \vx_1, &\quad \ve R = \frac{1}{2}(\vx_1+\vx_2), \nonumber\\
  \ve k_a = \frac{1}{2}(\vk_2 - \vk_1), &\quad \ve k_b = \vk_1 + \vk_2, \label{coordTrans}
 \end{align}
 in eq.~(\ref{mCFCLPT1}),
and perform Gaussian integrations in $\vk_a$ and $\vk_b$ (see e.g. appendix C of \cite{Carlson:2012bu}) to get 
the weighted correlation function in CLPT
\begin{align} \label{mCFCLPT2}
1+W(r) &= \frac{1}{\bar{m}^2}
\int  \frac{d^3 q \, e^{-\frac{1}{2}(\vr-\vq)^T\mathbf{A}_L^{-1}(\vr-\vq)} }{(2\pi)^{3/2}|\mathbf{A}_L|^{1/2}}
\int  \frac{d^3 Q \,e^{-\frac{1}{2}(\ve R-\ve Q)^T\mathbf{C}^{-1}(\ve R-\ve Q)}  }{(2\pi)^{3/2}|\mathbf{C}|^{1/2}} 
\Big(1+ \mathcal{I} \Big),
\end{align}
with matrix 
\begin{equation}
 C_{ij}(\vq) = \sigma_\Psi^2 \delta_{ij} - \frac{1}{4}A_{ij}^L(\vq).
\end{equation}
The complete expression for the function $\mathcal{I}(\vr,\vq,\ve R, \ve Q)$ is given in eq.~(\ref{I02}).
We notice that if $1+ \mathcal{I}$ is not a function of $\ve Q$, it
can be pulled out of the $\ve Q$ integral and due to that $C_{ij}$ depends only on $\vq$, 
the integration over $\ve Q$ gives 1. This is the case of the ``standard'' correlation function in
CLPT, where by statistical homogeneity we can shift all arguments of the fields inside the correlator in eq.~(\ref{mCFCLPT1}) 
by a vector $-\vq_1$ and, since the arguments $\vx_{1,2}$ are not present, we obtain 
$\mathcal{I}=\mathcal{I}(\vq)$, reducing the double Gaussian convolution in eq.~(\ref{mCFCLPT2}), to a single three-dimensional convolution.  
\begin{figure}
	\begin{center}
	\includegraphics[width=3 in]{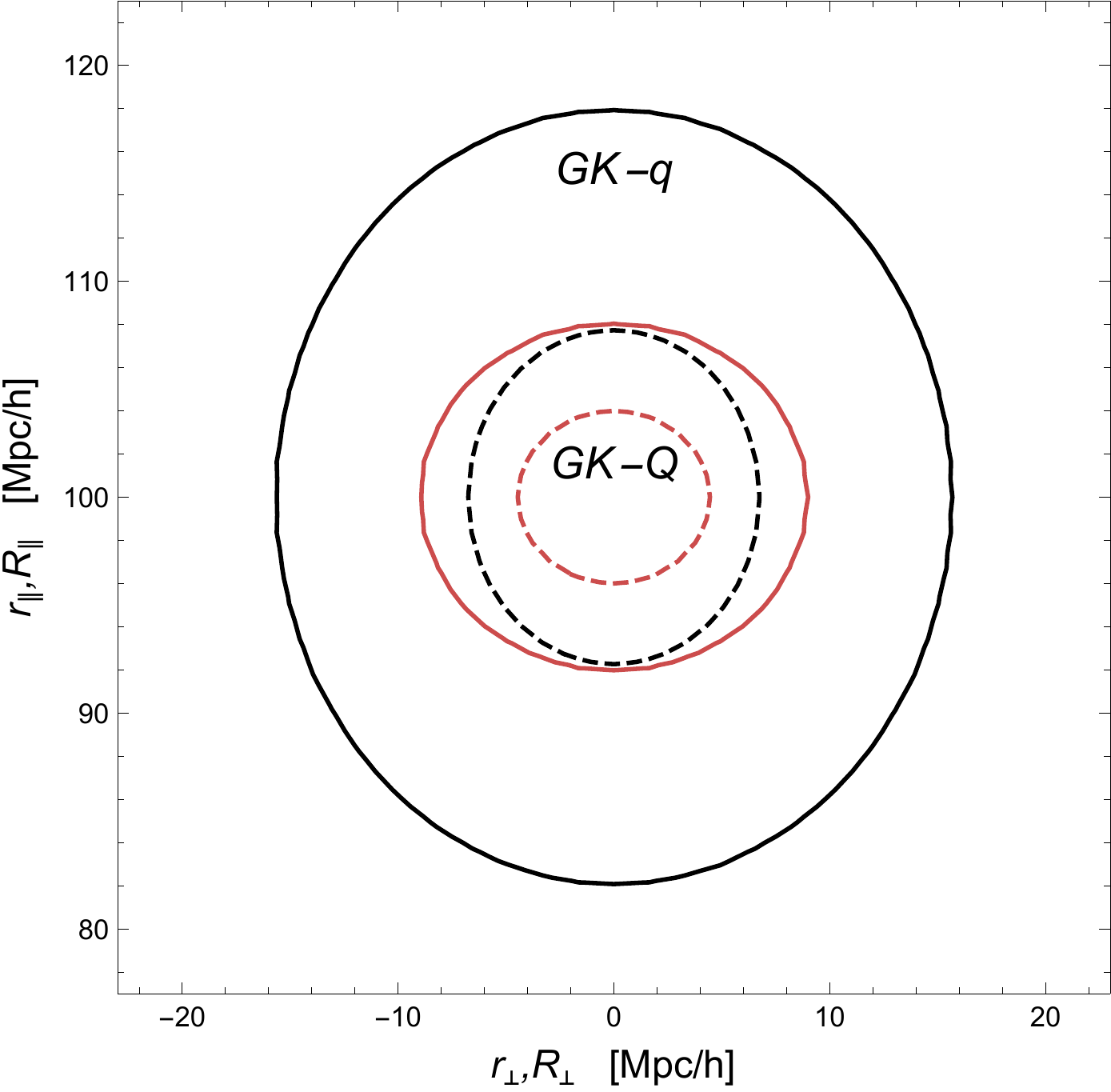}	
	\caption{Gaussian kernels in eq.~(\ref{mCFCLPT2}). {\it GK}-$q$ is the kernel of the $q$-integral (black curves) and 
	{\it GK}-$Q$ the kernel of the $Q$-integral (red curves). Dashed and solid curves show the regions that enclose the  68\% and 95\% of the volume
	respectively. A similar plot can be found in \cite{Tassev:2013rta}.
	\label{fig:GaussianKernels}}
	\end{center}
\end{figure}

In figure~\ref{fig:GaussianKernels} we show contour plots for the two Gaussian kernels appearing in eq.~(\ref{mCFCLPT2}). 
{\it GK}-$q$ is the kernel of the $q$-integral and 
{\it GK}-$Q$ the kernel of the $Q$-integral. These are plotted as a function of $r_\parallel$ ($R_\parallel$) and $r_\perp$ ($R_\perp$), 
the components of $\vr$ ($\ve R$) parallel and perpendicular
to the Lagrangian coordinate $\vq$ ($\ve Q$) with $Q=q=100\,\text{Mpc}/h$ fixed.\footnote{Reference \cite{Tassev:2013rta} shows that {\it GK}-$q$ gives the
probability of finding two dark matter particles separated by a distance $\vr$, given that they were separated by a distance $\vq$ at an earlier time.} 
We notice that both kernels have their maximum value at
$Q=q=100\,\text{Mpc}/h$, but the {\it GK}-$Q$ is more sharply peaked because the determinants of the correlation matrix comply with $|\mathbf{C}|<|\mathbf{A}_L|$. 
This observation suggests to replace the Gaussian kernel {\it GK}-$Q$ by a Dirac Delta function 
$\dD(\ve R - \ve Q)$, allowing us to perform the
$d^3Q$ integral to obtain
\begin{align}\label{preM}
\langle G_{1}F_{\vx,1}G_{2}F_{\vx,2} \rangle &= 
\int  \frac{d^3 q \, e^{-\frac{1}{2}(\vr-\vq)^T\mathbf{A}_L^{-1}(\vr-\vq)} }{(2\pi)^{3/2}|\mathbf{A}_L|^{1/2}}
\Big(1+ \mathcal{J}(\vr,\frac{1}{2}(\vr\pm\vq)) \Big),
\end{align}
with $\mathcal{J}$ given in eq.~(\ref{J0}).
Now, by noticing from figure~\ref{fig:GaussianKernels} that the largest contribution to the above integral is given by $\vq = \vr$, we replace the argument
$\frac{1}{2}(\vr\pm\vq)$ by $\vq$ or $\ve 0$, either if $\pm=+$ or $-$. 
Following these lines, in appendix~\ref{app:mCF} we arrive at
\begin{equation}\label{1pWM}
1+ W^\text{W16}(r) = \int \frac{d^3 q}{(2\pi)^{3/2}|A_L|^{1/2}} e^{-\frac{1}{2}(\vr-\vq)^T\mathbf{A}^{-1}_L(\vr-\vq)}
\Big(1+ J^{0}(\vq,\vr) \Big)
\end{equation}
where the label ``$0$'' in the $J$ function indicates that we have factorized zero-lag correlators $\sigma_R$ that are canceled out by the
square of the mean mark. To third order in fluctuations and bias expansion, this function is given by 
\begin{align}\label{J0main}
&1+J^0(\vq,\vr)=  1 - \frac{1}{2}A^\text{loop}_{ij}G_{ij} + \frac{1}{6} W_{ijk}\Gamma_{ijk} +
     b_1^2 \xi_L   - 2 b_1  U_i g_i - (b_2+b_1^2) U_iU_j G_{ij} \nonumber\\
 &\quad  - 2 b_1b_2\xi_L U_i g_i + B_1^2  \xi_{RR}  - 2B_1 U_i^R g_i- (B_2+B_1^2)U_i^{R} U_j^{R} G_{ij} - 2 B_1B_2 \xi_{RR} U^{R}_i g_i\nonumber\\
 &\quad  + 2b_1B_1 \xi_{R} - 4b_1B_1 U_i U_j^{R} G_{ij}  
        - 2b_1^2B_1\xi_L  U_i^{R} g_1 - 2B_1^2b_1\xi_{RR} U_i g_i \nonumber\\
 &\quad  - 2(b_2 + b_1^2)B_1 \xi_{R} U_i g_i -  2(B_2 + B_1^2)b_1  \xi_{R} U_i^{R} g_i   - b_1 A^{1000}_{ij}G_{ij} -B_1A^{0010}_{ij}G_{ij}  \nonumber\\
 &\quad  - b_2U^{2000}_ig_i  - B_2U^{0020}_ig_i   -  b_1^2 U^{1100}_ig_i - B_1^2 U^{0011}_ig_i   -2 B_1 b_1 (U^{1010}_i+U^{1001}_i)g_i. 
\end{align}
Hence, eqs.~(\ref{1pWM}) and (\ref{J0main})
give the higher order generalization of the linear expression found in \cite{White:2016yhs}; hereafter we will refer to this perturbative model as W16, and to the linear pieces as the Zeldovich Approximation (ZA) mCF.
In the above equation the $\vr$ dependence appears only through the tensors $g_i = (\mathbf{A}^{-1}_L)_{ij}(q_j-r_j)$, 
$G_{ij} = (\mathbf{A}^{-1}_L)_{ij} -g_ig_j$ and $\Gamma_{ijk}=(\mathbf{A}^{-1}_L)_{\{ij}g_{k\}} -g_ig_jg_k$, while $\xi$, $U$, $A$ and $W$ 
are functions of $\vq$ only.
$U(\vq)$ functions are defined as
\begin{equation}\label{Ufunct}
 U^{mnpq(r)}_i = \langle \delta^m(\vq_1)\delta^n(\vq_2)\delta^p_R(\vq_1)\delta^q_R(\vq_2) \Delta_i^{(r)}\rangle_c.
\end{equation}
The numbers $m,n,p,q$ denote powers, and $(r)$ is the perturbation theory order of the displacement fields. $U\equiv U^{1000}$ and
$U^R\equiv U^{0010}$.
The density fields are linear and are assumed Gaussian, thus only a few $U$ functions do not vanish.
Similarly we have defined
\begin{equation}\label{Afunct}
 A_{ij}^{mnpq(rs)} \equiv \langle \delta^m(\vq_1)\delta^n(\vq_2)\delta^p_R(\vq_1)\delta^q_R(\vq_2) \Delta_i^{(r)}\Delta_j^{(s)}\rangle_c,
\end{equation}
and  $A_{ij}^L = A_{ij}^{0000(11)}$, $ A_{ij}^\text{loop} = A_{ij}^{0000(22)} + 2 A_{ij}^{0000(13)}$.
Also, we have used the 3-point correlations 
\begin{equation}
 W_{ijk}=\langle \Delta_i^{(2)}\Delta_j^{(1)}\Delta_k^{(1)}\rangle_c + \text{cyclic perm.}
\end{equation}
Expressions for some of these $U$, $A$ and $W$ functions are known in the literature (see refs.~\cite{Carlson:2012bu,Vlah:2014nta} for $\Lambda$CDM and ref.~\cite{Aviles:2018saf} for MG), and in appendix \ref{app:kqfunctions} we show explicit expression for the rest.
The mCF is 
\begin{equation} \label{MmCF}
\mathcal{M}^\text{W16}(r;b_i,B_i) = \frac{1 + W^\text{W16}(r;b_i,B_i)}{1 + \xi^\text{CLPT}_X(r;b_i)}, 
\end{equation}
while
\begin{equation}\label{CLPTbtCF}
\xi^\text{CLPT}_X(r,b_1,b_2)=  W^\text{W16}(r;b_1,b_2,B_1=0,B_2=0)
\end{equation}
is the known biased tracers correlation function in CLPT \cite{Carlson:2012bu} with the resummation of ref.~\cite{Vlah:2015sea}.

We notice that the approach developed in this subsection and in appendix~\ref{app:mCF} to arrive to eq.~(\ref{1pWM}) is equivalent to substituting $\vx_{1,2}$ by $\vq_{1,2}$ in the arguments of the fields inside
the correlator of eq.~(\ref{mCFCLPT0}) from the beginning. For example, one of the contributions to the function $I$ that contains both Lagrangian ond Eulerian coordinates is $\xi_R(\vx_2-\vq_1)$. As long as the region over which the Gaussian kernels $GK$-$q$ and $GK$-$Q$ are non-negligible is smaller than the support of the smoothing kernel $W_R$, we can approximate $\xi_R(|\vx_2-\vq_1|) \simeq \xi_R(q) + \alpha (\sigma^2_\Psi/R^2)  \nabla^2\xi_R(q) + \cdots$, with $\alpha$ a constant depending on the particular spherically symmetric kernel $W_R$. If we drop terms of order $\sigma^2_\Psi/R^2$ from the contributions to the $I$ function we arrive to eq.~(\ref{1pWM}).

However, a simple substitution of $\lambda \delta (\vq_1) + \Lambda \delta_R(\vx_1) \rightarrow \lambda \delta (\vq_1) + \Lambda \delta_R(\vq_1) $ in the integrand of eq.~(\ref{mCFCLPT0}), is not justified even for linear density fields because the smoothed procedure is non local, and all matter particles at $\vx_a$ within the region $|\vx_a - \vx_1|<R$ contribute. So, the above substitution requires that about the same particles contribute to the smoothing in the initial slice, $|\vq_a - \vq_1|<R$. Hence in addition to the linear evolution of $\delta_R$ we have to impose that $|\Psi_a-\Psi_1|\sim \sigma_\Psi < R$. Clearly, to have a linear evolution for $\delta_R$ requires a large smoothing scale $R$, but the latter condition is independent and in general more restrictive.

In the next subsection, we will discuss a different approximation to handle the double convolution without using the substitution of $\vx_{1,2}$ by $\vq_{1,2}$.

\end{subsection}

\begin{subsection}{Approximating the CLPT marked correlation function}
\label{subsec:mCFCLPTapprox}

Solving numerically the double convolution in eq.~(\ref{mCFCLPT2}) is challenging and we do not attempt it in this work. Instead, in this subsection we put forward an approximation method that splits the function $I$ in eq.~(\ref{mCFCLPT1}) into pieces containing $\vq$, pieces containing combinations of $\vq_i$ and $\vx_j$, and pieces containing $\vr=\vx_2-\vx_1$:
\begin{equation}
 I=I_r(r) + I_q(q) + I_{r,q}(r,q).
\end{equation}
The first and second functions can be integrated by the standard methods of CLPT. For example, the term $\xi_{RR}(r) \in I_r$ yields
$B_1^2 \xi_{RR}(r)(1+\xi_\text{ZA}(r)) \in 1+ W$. The function $I_q(q)$ gives $1+\xi^\text{CLPT}_X(r)$. 
For the terms containing both $r_i$ and $q_i$ in their arguments we approximate
\begin{equation} \label{approxTW}
e^{\frac{1}{2}k_1^ik_2^jA^L_{ij}(\vq)} \approx 1+ \frac{1}{2}k_1^ik_2^jA^L_{ij}(\vq),    
\end{equation}
and solve
\begin{align} \label{mCFCLPTForapp1}
\langle G_{1}F_{\vx,1}G_{2}F_{\vx,2} \rangle\big|_{r,q} &=
\int \Dk{k_1}\Dk{k_2} \int d^3q_1 d^3q_2 
 e^{i\vk_1\cdot (\vx_1-\vq_1)}e^{i\vk_2\cdot (\vx_2-\vq_2)} e^{-\frac{1}{2}(\vk_1+\vk_2)^2\sigma^2_\Psi} \nonumber\\
 &\qquad \times I_{r,q} \left[ 1 +  \frac{1}{2}k_1^ik_2^jA^L_{ij}(\vq) \right].
\end{align}
This approximation allows us to write $1+W$ as a sum of Fourier transforms of known scalar functions constructed
out of 2- and 3-point correlation functions of Lagrangian displacements. We obtain (see appendix \ref{app:mCFapprox})
\begin{align}\label{1pWTW}
& 1 + W = 1 + \xi^\text{CLPT}_X(r) +  B_1^2  \xi_{RR}(r)(1 + \xi_\text{ZA}(r)) + 2b_1B_1 \bar{\text{x}}_{\xi_R}(r) 
+ 2 B_1 \bar{\text{x}}_{U_R}(r) \nonumber\\
& \quad +(B_2+B_1^2)  \bar{\text{x}}_{U_RU_R} 
  + 2 B_1B_2 \xi_{RR}(r) \bar{\text{x}}_{U_R}(r) + 4b_1B_1 \bar{\text{x}}_{U U_R}(r) 
        + 2b_1^2B_1 \bar{\text{x}}_{\xi U_R}(r)  \nonumber\\
& \quad       + 2B_1^2b_1 \xi_{RR}(r) \text{x}_{U}(r) +2(b_2 + b_1^2)B_1 \bar{\text{x}}_{\xi_R U}(r) +  2(B_2 + B_1^2)b_1  \bar{\text{x}}_{\xi_R U_R}(r) \nonumber\\
 &  \quad   +B_1  \bar{\text{x}}_{A^{0010}}(r)  + B_2 \bar{\text{x}}_{U^{0020}}(r)  
 +   B_1^2 \bar{\text{x}}_{U^{0011}}(r)   +2 B_1 b_1 \bar{\text{x}}_{U^{0110}}(r)  +2 B_1 b_1 \bar{\text{x}}_{U^{1010}}(r).
\end{align}
The dominant $\bar{\text{x}}$ functions above are
\begin{align}
\bar{\text{x}}_{\xi_R}(r) &= \xi_{R\sigma^2_\Psi}(r)  + \xi_{R\sigma^2_\Psi}(r)\xi_L(r) 
 - \xi^{[1,1]}_{R\sigma^2_\Psi}(r)\xi_L^{[1,-1]}(r), \\
\bar{\text{x}}_{U_R}(r) &= \xi_{R\sigma^2_\Psi}(r) + \xi^{\rm loop}_{R\sigma^2_\Psi}(r) + \frac{4}{3}\xi_{R\sigma^2_\Psi}(r)\xi_L(r) ,
 + \frac{2}{3}\xi^{[2,0]}_{R\sigma^2_\Psi}(r)\xi_L^{[2,0]}(r) \nonumber\\
 &\quad- \xi^{[1,1]}_{R\sigma^2_\Psi}(r)\xi_L^{[1,-1]}(r)-\xi^{[1,-1]}_{R\sigma^2_\Psi}(r)\xi_L^{[1,1]}(r), \label{barxUR}
\end{align}
where 
\begin{equation}
 \xi^{[m,n]}_{R\sigma^2_\Psi}(r) = \frac{1}{2\pi^2}\int_0^\infty dp\, p^{2+m} e^{-p^2\sigma^2_\Psi/2}\tilde{W}_R(p) P_L(p) j_n(pr), 
\end{equation}
are generalizations of the correlation function (see e.g. \cite{Schmittfull:2016jsw,Slepian:2016weg}). The rest of 
the $\bar{\text{x}}$ functions are given in appendix~\ref{app:mCFapprox}. Equivalently to eq.~(\ref{MmCF}), the mCF
 is 
 \begin{equation} \label{TWmCF}
\mathcal{M}(r;b_i,B_i) = \frac{1 + W(r;b_i,B_i)}{1 + \xi^\text{CLPT}_X(r;b_i)}. 
\end{equation} 

\begin{figure}
	\begin{center}
	\includegraphics[width=3.2 in]{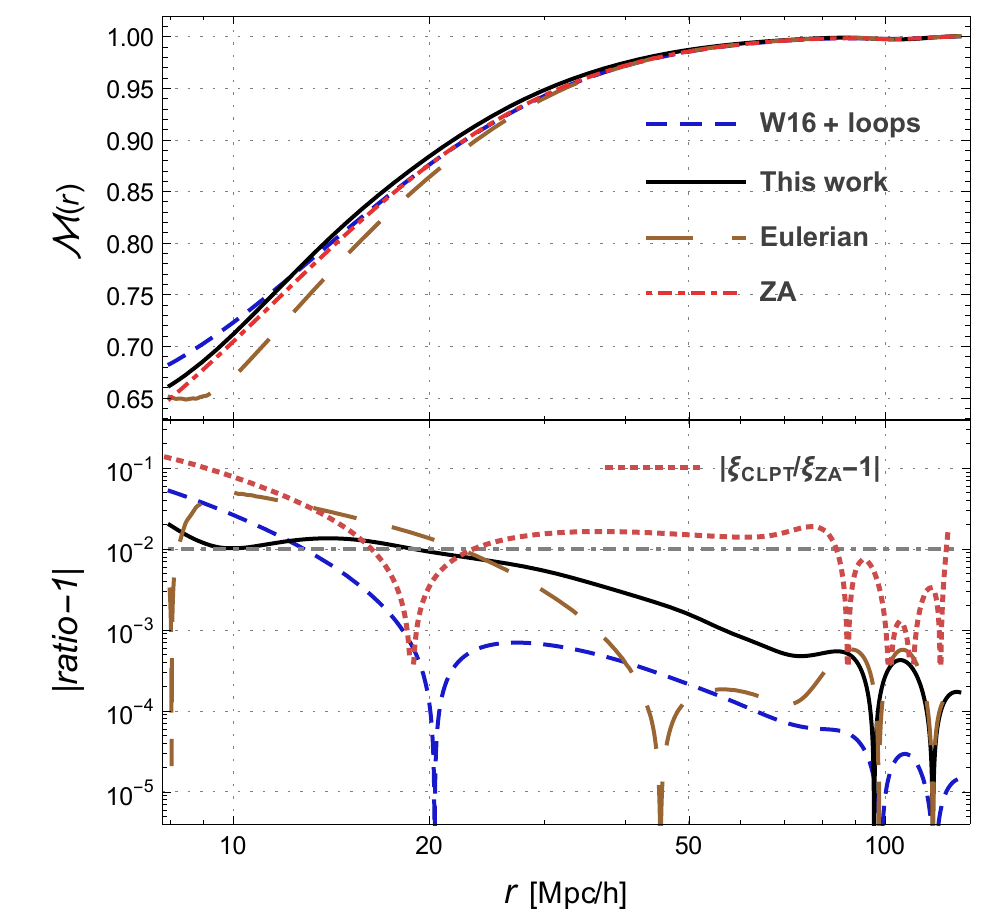}	
	\caption{Comparison of perturbative models of sections~\ref{sec:Eulerian}, \ref{subsec:mCFCLPT}, and \ref{subsec:mCFCLPTapprox}. 
	The upper panel shows the marked correlation functions with parameters 
	$b_1=1.2$, $b_2=0.2$, $B_1=-0.7$, $B_2=B_1^2$, for the different methods:
	solid black curve is the one presented in this subsection [eq.~(\ref{1pWTW})]; dashed blue the method of  eq.~(\ref{1pWM}), which is the method introduced in ref.~\cite{White:2016yhs} plus 1-loop  contributions; in dot-dashed red, the ZA; and in long-dashed brown, the Eulerian linear model of eq.~(\ref{mCFE}). The lower panel shows the relative differences with respect to the ZA.  
	The gray dot-dashed horizontal line denotes the $1\%$ differences. The red dotted curve in the lower panel shows the relative differences between the CLPT 1-loop and ZA standard correlation functions.
	\label{fig:CompareModels}}
	\end{center}
\end{figure}

Figure \ref{fig:CompareModels} shows the mCF using different analytical methods for the MG Hu-Sawicki F5 model: solid black curve is the one presented in this subsection (``This Work'', from now on); dashed blue the method of  eq.~(\ref{1pWM}), which is that introduced in ref.~\cite{White:2016yhs} plus 1-loop contributions; in dot-dashed red, the Zeldovich Approximation, 
which in this work means the linear model of ref.~\cite{White:2016yhs}; and, in long-dashed brown, the Eulerian linear model of eq.~(\ref{mCFE}).
We are using Lagrangian local biases $b_1=1.2$ and $b_2=0.2$, and mark parameters $B_1=-0.7$, $B_2=B_1^2$. In the lower panel we show the relative difference between the models and the ZA. The  agreement between them is very good, even for the linear Eulerian theory, below 1\% for scales above $\sim 20 \,\text{Mpc}/h$. The relative differences between the CLPT standard correlation function with and without loop contributions is shown in the lower panel of figure~\ref{fig:CompareModels} (red dotted curve); by comparing it with the dashed blue curve (which shows the relative differences of the mCF W16+1-loop with the ZA) we confirm that mCFs which efficiently enhance low density regions in the sky are indeed more linear than the standard correlation function.

To summarise, ``This Work'' method is a mixture of Lagrangian and Eulerian 
schemes that splits the function $I$ according to its arguments. The pieces depending only on coordinates $\vq$ and on
coordinates $\vx$ are treated exactly (within CLPT), and hence more appealing
than the method of \cite{White:2016yhs}, which relies on the substitution of $\vx_{1,2}$ by $\vq_{1,2}$.  Meanwhile, the $I_{r,q}$ piece relies on the approximation of eq.~(\ref{approxTW}) which is not well justified since
$\frac{1}{2} k_1^ik_2^jA^L_{ij}(\vq\rightarrow \infty)\rightarrow \sigma^2_\Psi k_1 k_2 (\hat{\vk_1}\cdot \hat{\vk_2})$,
and $\sigma^2_\Psi$ is not small. However, this is not as catastrophic as it appears because the oscillatory nature of the integrand in eq.~(\ref{mCFCLPTForapp1}) gives much more weight to low values of $k_1$ and $k_2$. On the other hand, a mixture of Lagrangian and Eulerian 
schemes has a disadvantage in describing the BAO peaks in the correlation function. In section \ref{sec:CS}, we will verify various approaches to model the marked correlation function shown in figure~\ref{fig:CompareModels} using the measurements from {\it N}-body simulations.

\end{subsection}

\end{section}

\begin{section}{Degeneracies between the mark and the bias}\label{sect:Degeneracies}

As discussed above, a mark that enhances low-density regions typically has $B_1<0$, meaning that these terms contribute by lowering the mCF, 
the smaller $B_1$ the larger the effect. In the left panel of figure~\ref{fig:mCFmatter} 
we show this effect on the mCF for matter  ($b_i=0$) for the gravitational models GR, F6, F5 and F4. 
We notice that for models that depart more from $\Lambda$CDM, the effect is more pronounced, which is the consequence of a enhanced clustering in $f(R)$ than in GR. This behavior has been observed recently in simulations \cite{Valogiannis:2017yxm}.

In the right panel of figure~\ref{fig:mCFmatter}, we plot the mCF for biased tracers with linear and second order biases  
$b_1=-0.3,\,0,\,0.5\,,1.5$,  $b_2=0$, and mark parameters fixed to
$B_1=-1$, $B_2=0$. Contrary to what 
may be naively expected, the effect of a greater large-scale bias is similar to making $B_1$ more negative, bringing down even more the mCF. This is because larger bias increases functions $W$ and $\xi_X$  of the mCF, while $B_1$ only affects $W$.

\begin{figure}
        \begin{center}
	\includegraphics[width=2.8 in]{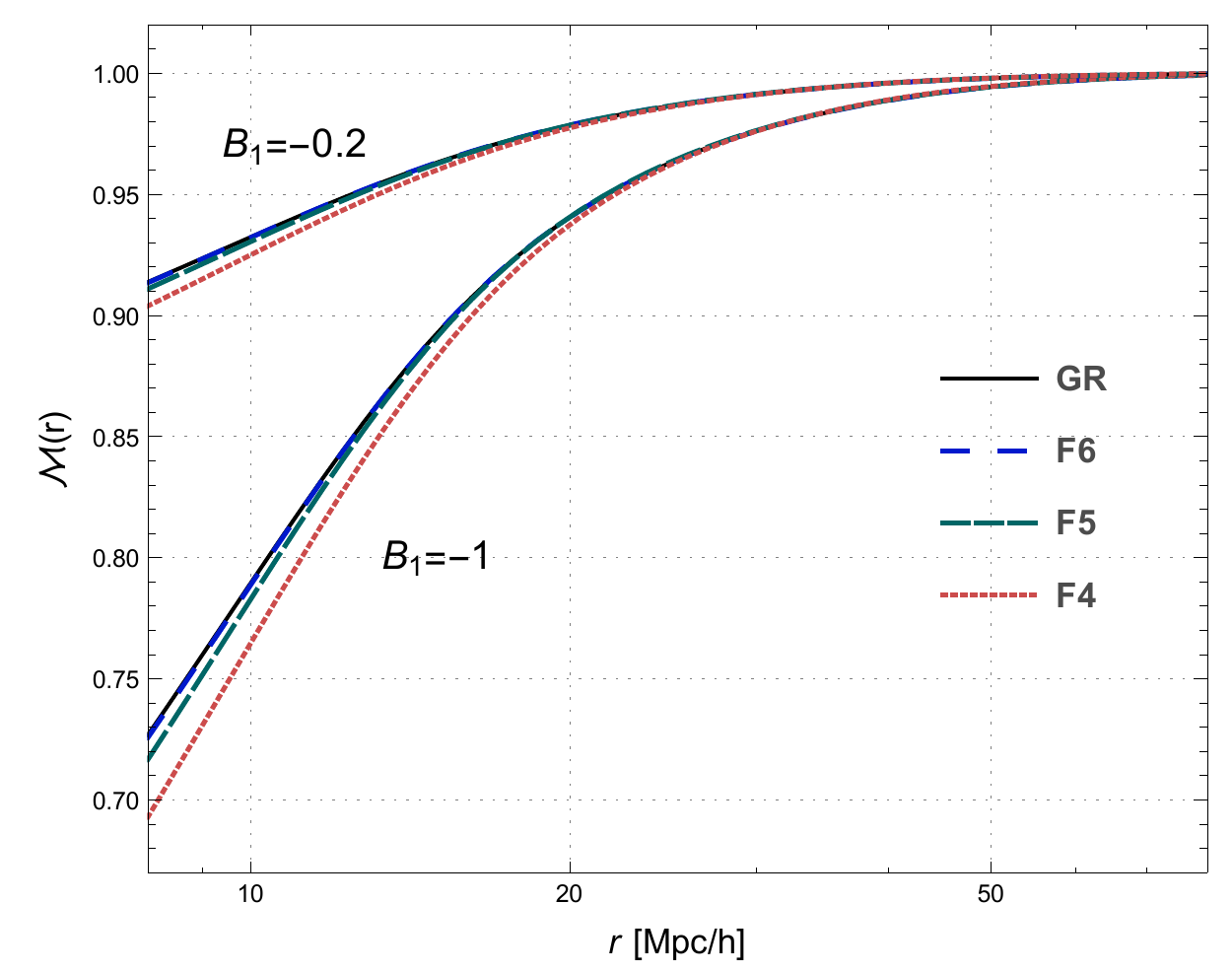}	
	\includegraphics[width=2.8 in]{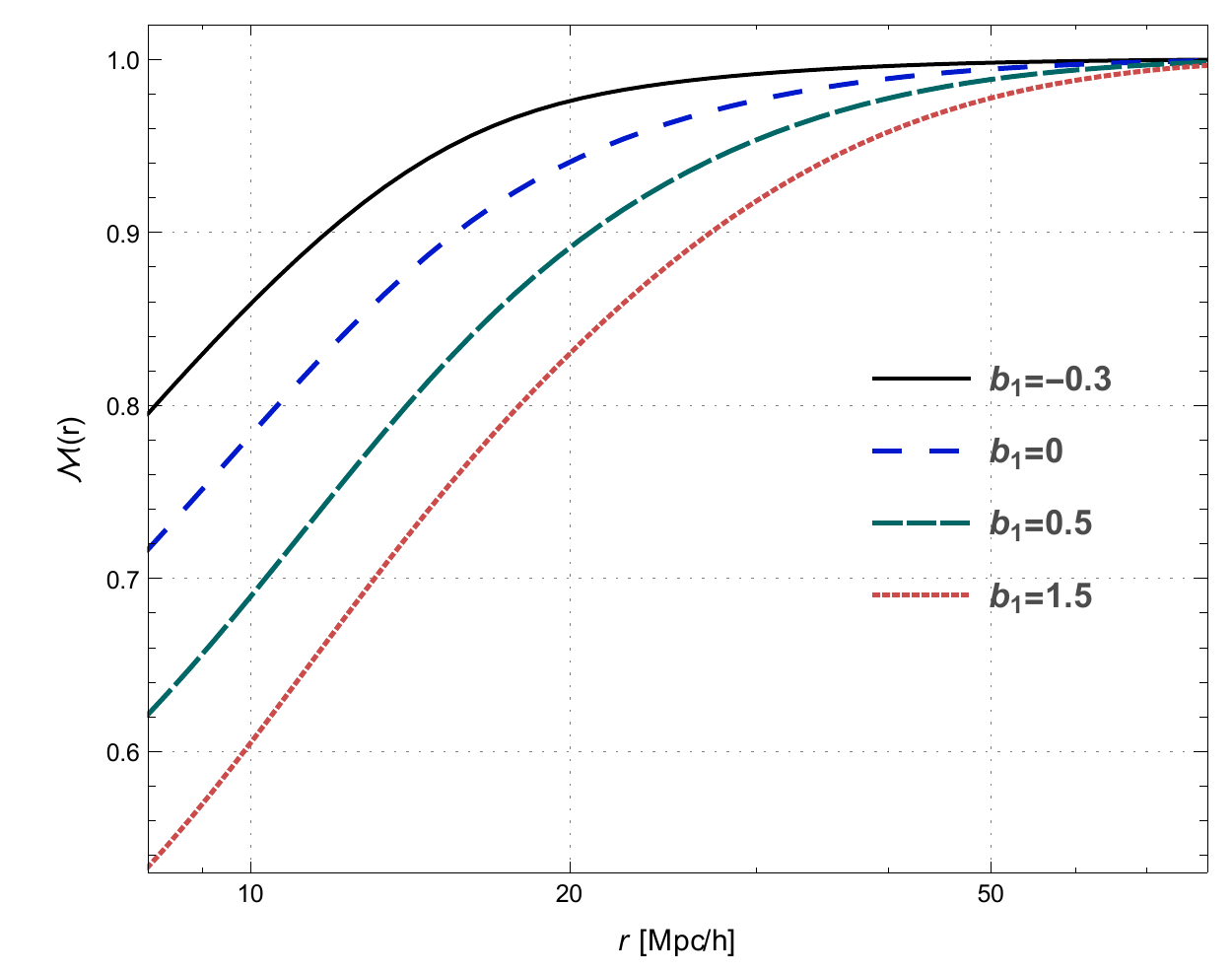}	
	\caption{Marked correlation function for matter for GR, F6, F5 and F4 models.
	In the left panel we consider matter ($b_1=b_2=0$) and set $B_1=-1$, $B_2=0$ (lower curves) and $B_1=-0.2$, $B_2=0$ (upper curves).  
	The right panel uses F5 model with fixed $B_1=-1$, $B_2=0$, $b_2=0$ and  show the results
	with Lagrangian local biases $b_1=-0.3,\,0,\,0.5\,,1.5$ from top to bottom. 
	\label{fig:mCFmatter}}
	\end{center}
\end{figure}
\begin{figure}
        \begin{center}
	\includegraphics[width=2.8 in]{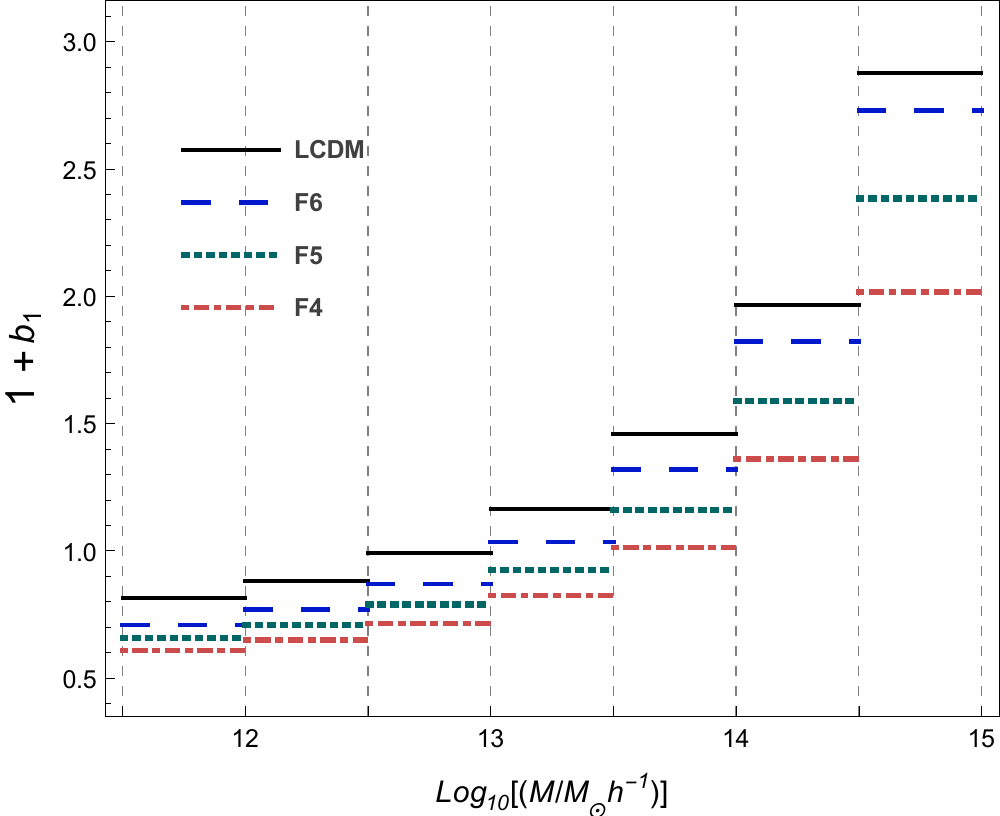}
	\includegraphics[width=2.8 in]{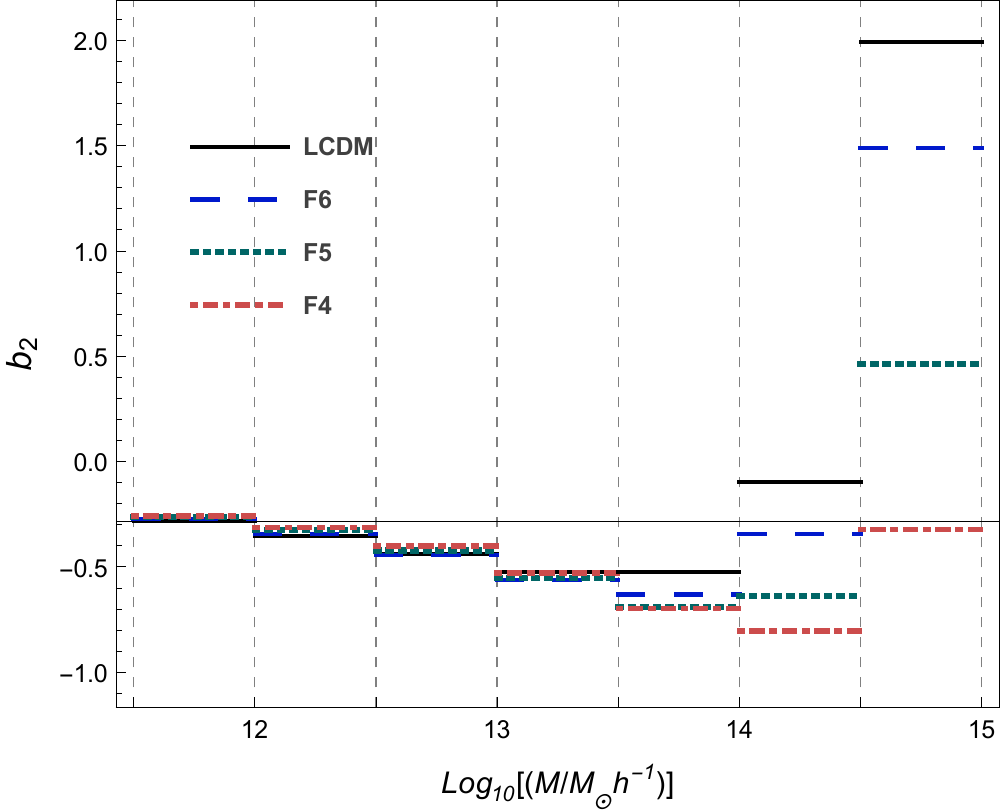}			
	\caption{Large scale bias $b_\text{LS}=1+b_1$ and second order local Lagrangian bias $b_2$ binned over halo mass intervals, for GR, F6, F5 and F4 models. 
	\label{fig:bias}}
	\end{center}
\end{figure}

\begin{figure}
        \begin{center}
	\includegraphics[width=3 in]{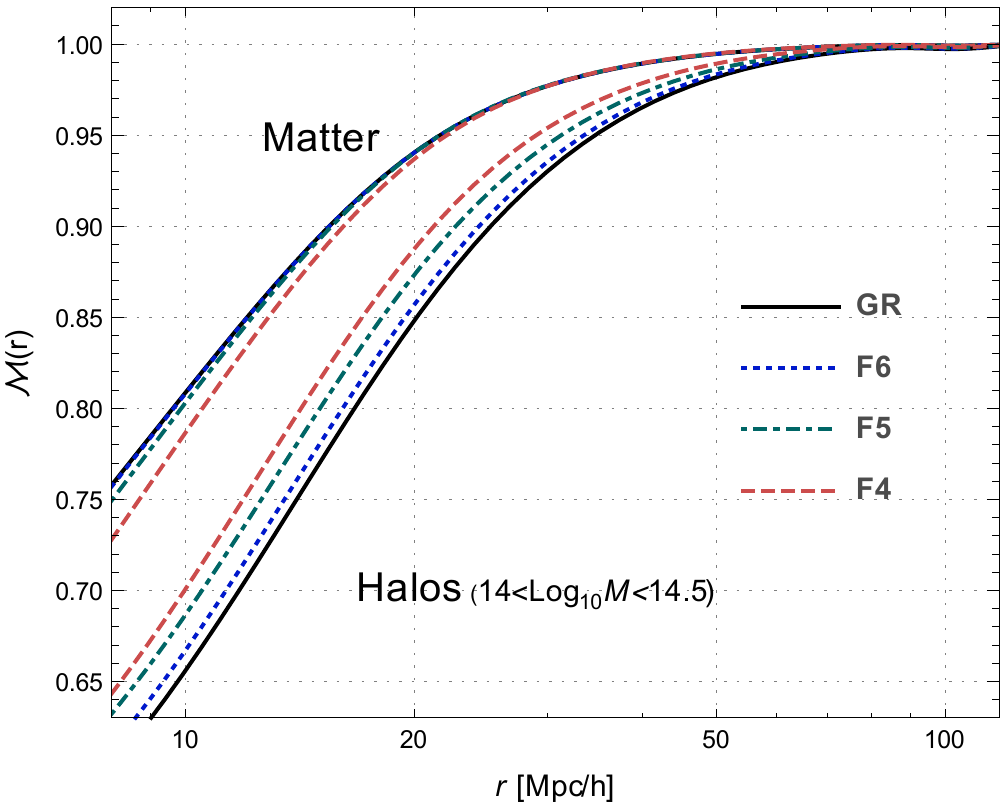}		
	\caption{Marked correlation function for matter and tracers for models $\Lambda$CDM, F6, F5 and F4 at redshift $z=0$.
	We fix $B_1=-1$, $B_2=0$ and take the bias values from figure~\ref{fig:bias} for the interval
	$10^{14}<M/M_\odot<10^{14.5}$: $b_1^\text{GR}=1.15$, $b_1^\text{F6}=1$, $b_1^\text{F5}=0.75$, $b_1^\text{F4}=0.5$. The upper curves
	are for matter and the lower curves for halos.
	\label{fig:mCFmattertracers}}
	\end{center}
\end{figure}

On the other hand, it has been shown that in MG models that produce more clustering the linear local bias parameters have smaller values than those obtained in $\Lambda$CDM \cite{Arnold:2018nmv,Aviles:2018saf}. 
This can be understood because typically local biases depend on two parameters: 
the variance of linear fluctuations in a given region of Lagrangian radius $R_*$,
$\sigma^2(M)$, and the density threshold for collapse $\delta_c(M)$, with $M$ the mass enclosed by a spherical region of radius $R_*$. 
The former is bigger in MG than in GR because there is a major clustering, $P_L^\text{MG}>P_L^\text{GR}$, 
while the latter is smaller in MG because of the extra attractive fifth-force; note also that a
violation of Birkhoff theorem in MG implies the density threshold for spherical collapse becomes mass dependent. Schematically,
these two quantities appear in the combination $\nu_c= \delta_c(M)/\sigma^2(M)$, hence one expects to obtain $b^\text{MG}<b^\text{GR}$. 
In figure~\ref{fig:bias}
we show the linear and second order local biases for halos obtained with the Sheth-Tormen like prescription 
of \cite{Aviles:2018saf} and averaged over mass bins as in refs.~\cite{Matsubara:2008wx,Valogiannis:2019xed}. 
We interpret figure~\ref{fig:bias} as a more rapid relaxation of linear bias in MG than in GR due to a more efficient
formation of massive halos.


Although this method to obtain biases in chameleon MG theories from the peak-background-split prescription gives reasonable values for the local bias parameters \cite{Aviles:2018saf,Valogiannis:2019xed}, we will find later in section~\ref{sec:CS} that it is not sufficiently accurate for our purposes on the marked correlation function, and hence we will be forced to 
fit the biases directly from the simulations. Therefore, figure~\ref{fig:bias} should be taken mainly as indicative, to gain intuition on the behaviour of the relative bias values in different gravitational theories.

In figure~\ref{fig:mCFmattertracers} we show plots for matter and halos with masses in the interval $10^{14}<M/M_\odot<10^{14.5}$. In this case we 
find that while for matter MG mCFs lie below that of GR, for tracers the opposite happens, the halos MG mCFs are brought above that of GR. This effect of inversion of the trends for the mCFs for tracers and matter can be interpreted by considering the mean mark, $\bar{m} \approx (1+b_1)B_1 \sigma_R^2$, which shows
that for the unbiased case  $\bar{m}^\text{MG}_\text{matter} < \bar{m}^\text{GR}_\text{matter}$, because $\sigma_R^\text{MG} > \sigma_R^\text{GR}$ and $B_1$ is negative. However, if the differences in linear local bias are sufficiently large they yield  $\bar{m}^\text{MG}_\text{tracers} > \bar{m}^\text{GR}_\text{tracers}$.

\end{section}
 
\begin{section}{Comparing to simulations}
\label{sec:CS}

The simulation suite used in this paper is the Extended LEnsing PHysics using ANalaytic ray Tracing ({\sc elephant}) simulations \cite{2018MNRAS.476.3195C} run with the \texttt{ECOSMOG} code \cite{2012JCAP...01..051L}. This suite contains five realisations of the initial conditions and for each realisation we have one simulation of $\Lambda$CDM (GR) together with three simulations of the Hu-Sawicki $f(R)$ model with parameters $f_{R0} = -10^{-4}$ (F4), $f_{R0} = -10^{-5}$ (F5) and $f_{R0} = -10^{-6}$ (F6). It also contains galaxy mock catalogs that were made with a Halo Occupation Distribution 
method. The HOD parameters for the $\Lambda$CDM model are the best-fit parameter values from the CMASS data \cite{2013MNRAS.428.1036M}. For the $f(R)$ models, we tune the HOD parameters so that we reproduce the correlation function in the $\Lambda$CDM model. The simulations were run in a box of size $L=1024$ Mpc$/h$ with $N = 1024^3$ particles and the cosmological parameters used to make the initial conditions were $\Omega_b = 0.046$, $\Omega_\Lambda = 0.719$, $\Omega_m = 0.281$, $h = 0.697$, $\sigma_8 = 0.82$ and $n_s = 0.971$, and for our analysis we choose snapshots at redshift $z=0.5$ 

To compute the mark for each of our tracers (dark matter particles, halos or mock galaxies) we binned the particles/halos/mock galaxies to a grid with gridsize of $20$ Mpc$/h$ (corresponding to a $N = 52^3$ grid) using a Nearest Grid Point (NGP) assignment scheme to get an estimate for the density for which the mark depends on.

The correlation functions were computed using the Correlation Utilities and Two-point Estimates (\texttt{CUTE}) code\footnote{The code can be found at https://github.com/damonge/CUTE} \cite{2012arXiv1210.1833A}. We computed both standard ($\xi(r)$) and weighted ($W(r)$) two-point correlation functions, the latter using the mark as a weight in \texttt{CUTE}. From this the marked correlation function follows simply as $\mathcal{M}(r) = \frac{1+W(r)}{1+\xi(r)}$. 

We consider the White-mark with $\rho_*=10$, $p=7$, corresponding 
to coefficients $C_0=1$, $C_1=-0.64$, $C_2=0.46$ in the 
Taylor expansion of the mark function in eq.~(\ref{Taylorm}). In our analytical models
we smooth the matter fields that assign the mark with a top-hat filter $W_R$ of radius $R=10\, \text{Mpc}/h$. 
 
\begin{subsection}{Matter particles}\label{subsec:CSmatter}

\begin{figure}
        \begin{center}
	\includegraphics[width=3 in]{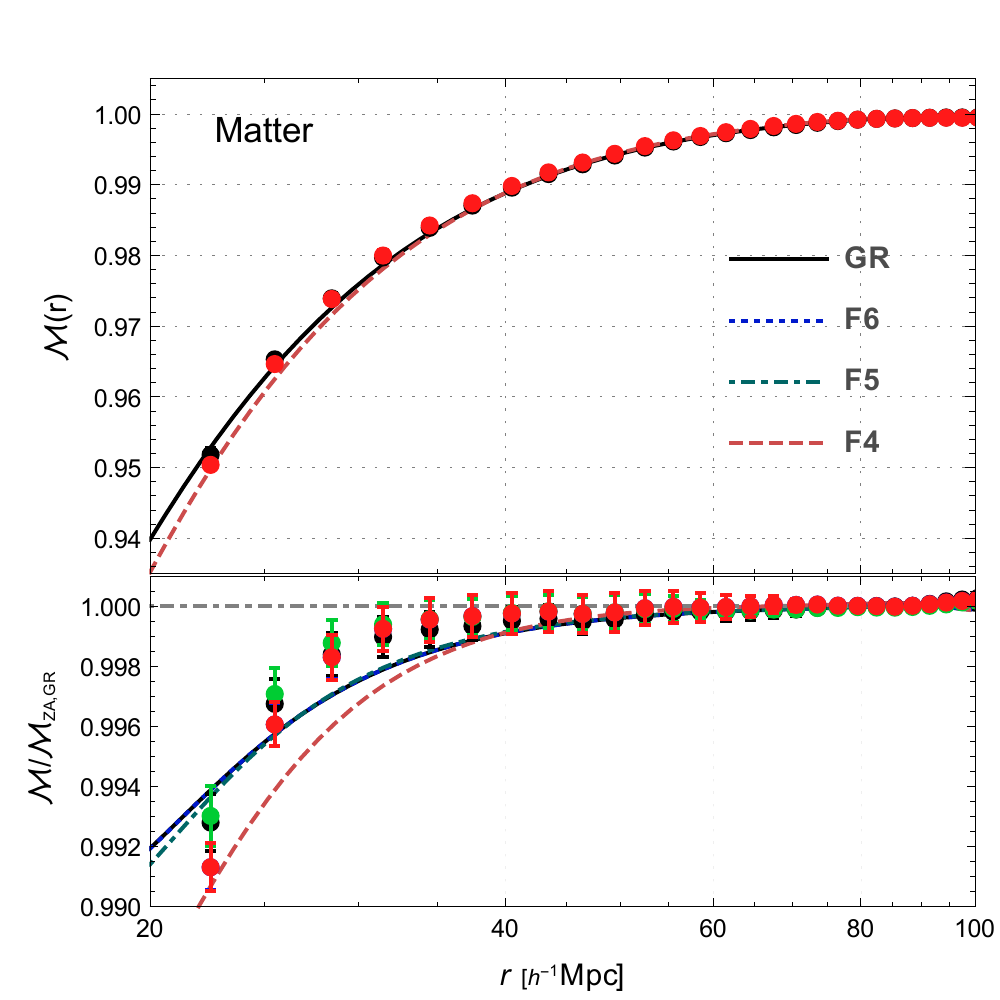}
	\caption{The top panel shows the marked correlation functions for matter particles in GR and F4 HS gravity and the bottom panel F4, F5, F6, and GR ratios to the ZA in GR.
	\label{fig:matter_mCF}}
	\end{center}
\end{figure}
\begin{figure}
    \begin{center}
	\includegraphics[width=5.4 in]{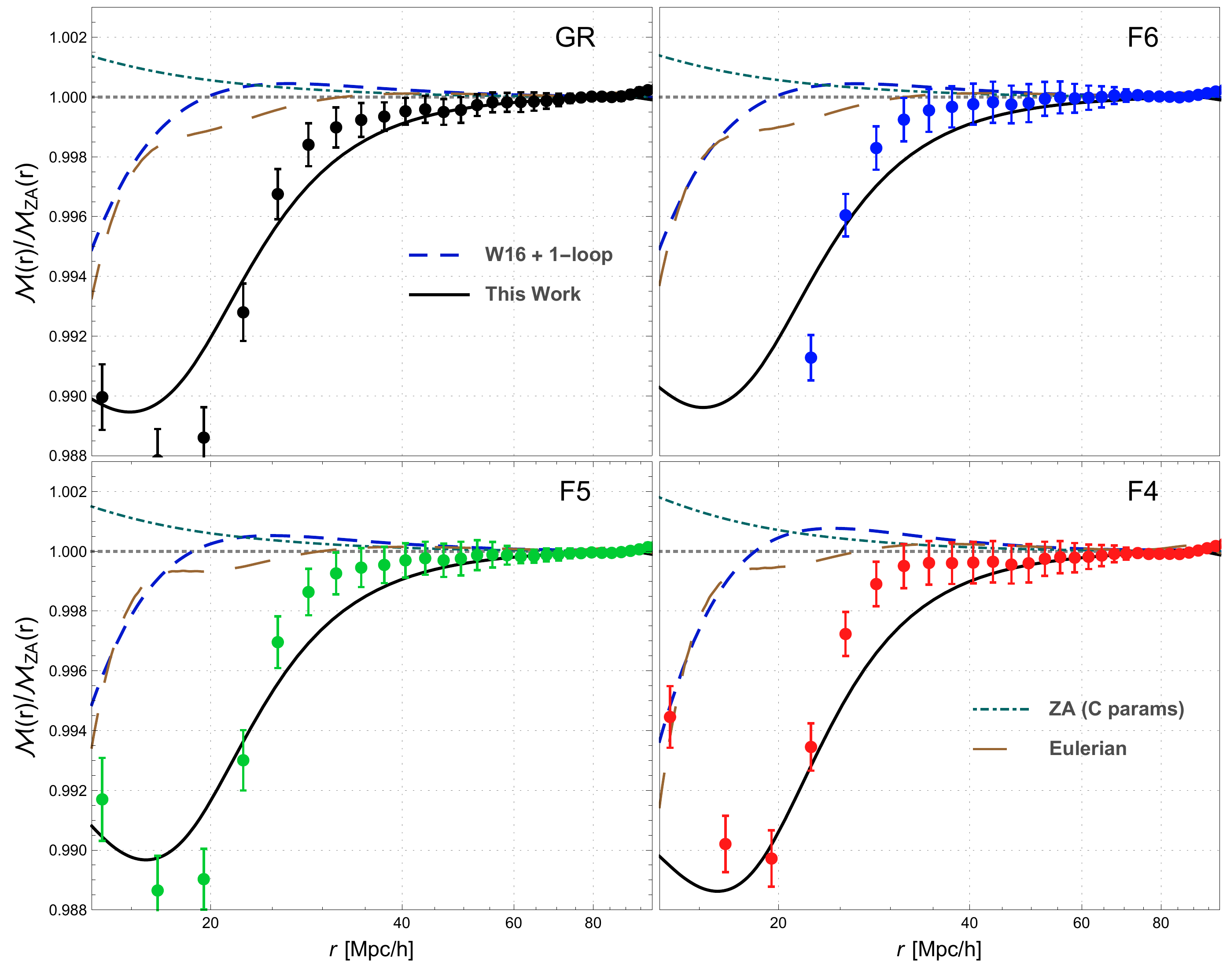}		
	\caption{Ratios of different mCFs analytical models to the ZA mCF for 
	GR, F6, F5 and F4. We show the W16 model plus 1-loop corrections of eq.~(\ref{MmCF}) (dashed blue); 
	the model of this work [eqs.~(\ref{1pWTW},\ref{TWmCF})] (solid black); the linear Eulerian model of eq.~(\ref{mCFE}) (long-dashed brown);
	and the ZA but using the Taylor expansion parameters of the mark function $C$ in eq.~(\ref{Taylorm}) (dot-dashed green), 
	instead of the resummed expansion parameters $B$.   
	\label{fig:mattermCFsratios}}
	\end{center}
\end{figure}

In this subsection we confront different analytical methods against matter data, since in this case the mCF does not suffer from biasing-marking degeneracies and we can observe more neatly the effects of the weights. 
The expansion parameters $B$ are obtained from eq.~(\ref{BnRenorm}): $B_1^\text{GR}=-0.6495 $, $B_1^\text{F6}=-0.6497$, 
$B_1^\text{F5}=-0.6505$, $B_1^\text{F4}=-0.6522$, $B_2^\text{GR}= 0.4628$, $B_2^\text{F6}=0.4502$, $B_2^\text{F5}=0.4495$, $B_2^\text{F4}=0.4480$; which are numerically very close to $C_1$ and $C_2$ values because $C_0=1$ and the variances $\sigma^2_{RR}$ are small.

In the top panel of figure~\ref{fig:matter_mCF} we show the analytical mCFs for GR and F4 obtained 
with the perturbative method presented in section~\ref{subsec:mCFCLPTapprox} [eqs.~(\ref{1pWTW}, \ref{TWmCF})], showing that they follow 
reasonably well the trends of the matter mCF extracted from simulations. The bars correspond to the root-mean-squared (RMS) error 
of the five simulation boxes available for each model.
In the bottom panel, we show the ratios of the mCF for each model to the Zel'dovich (ZA) model in GR -- by ZA we mean the original model in ref.~\cite{White:2016yhs}, that is, the linear piece of eq.~(\ref{MmCF}). We notice that the data
for different gravitational theories are very close 
to each other, and the RMS errors, although 
small, are large enough such that one cannot distinguish between the models.
Hence, PT may not help to differentiate between gravitational theories with matter mCFs. One may try 
other values for parameters $\rho_*$ and $p$ in the White-mark, but as was shown in \cite{Valogiannis:2017yxm}, the differences of mCFs between different models remain almost the same. In the upcoming subsections we will see that this situation changes drastically by using tracers instead of matter. 
Nevertheless, the matter mCF data 
still enables us to compare the different perturbative methods, which is the main objective of this section. 

In figure~\ref{fig:mattermCFsratios} we show the ratios of the different analytical methods to the ZA model. On each panel
the dashed blue curves show the results with the W16 model plus 1-loop corrections of eq.~(\ref{MmCF}); 
solid black, the method presented in this work [eqs.~(\ref{1pWTW},\ref{TWmCF})] (``This Work''); long-dashed brown, the linear Eulerian model of eq.~(\ref{mCFE}); 
and dot-dashed green, the ZA model, but instead of using the resummed $B$ expansion parameters we use the Taylor coefficients $C$.  
The differences between the analytical methods are apparent but small, being lesser than the $1\%$. 
At scales $r>40\,\text{Mpc}/h$, all analytic models are indistinguishable and they are within the errors of the simulation data. 
At smaller scales, the method ``This Work'' 
outperforms the other perturbative approaches and captures pretty well the trend of the data all the way up to the smoothing scale 
$R=10 \,\text{Mpc}/h$, but it is not accurate enough to lie inside the RMS errors of the data. 

\end{subsection}

\begin{subsection}{Halos and mock galaxies}\label{subsec:CShalos}

\begin{figure}
        \begin{center}
	\includegraphics[width=2.8 in]{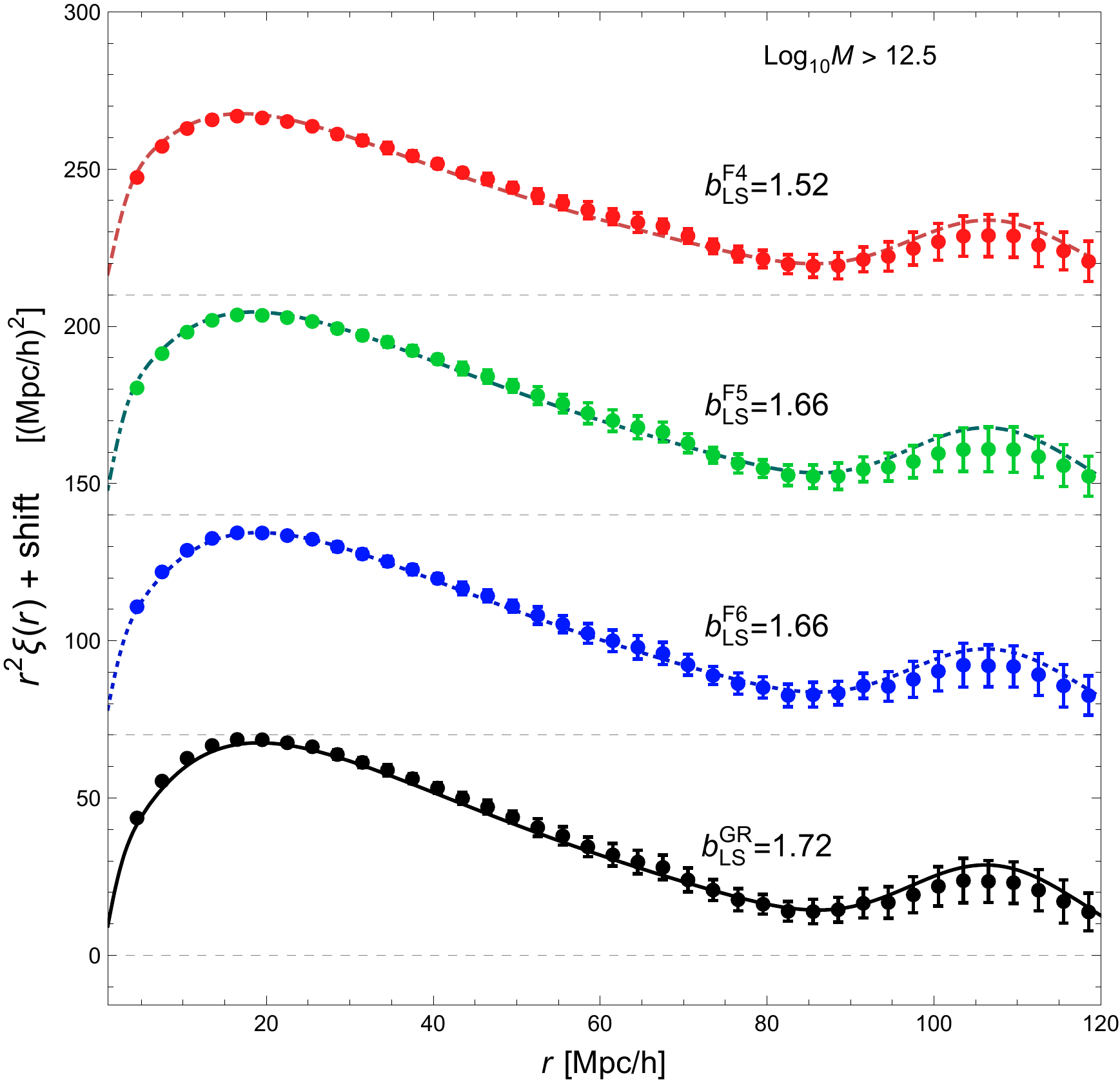}	
	\includegraphics[width=2.8 in]{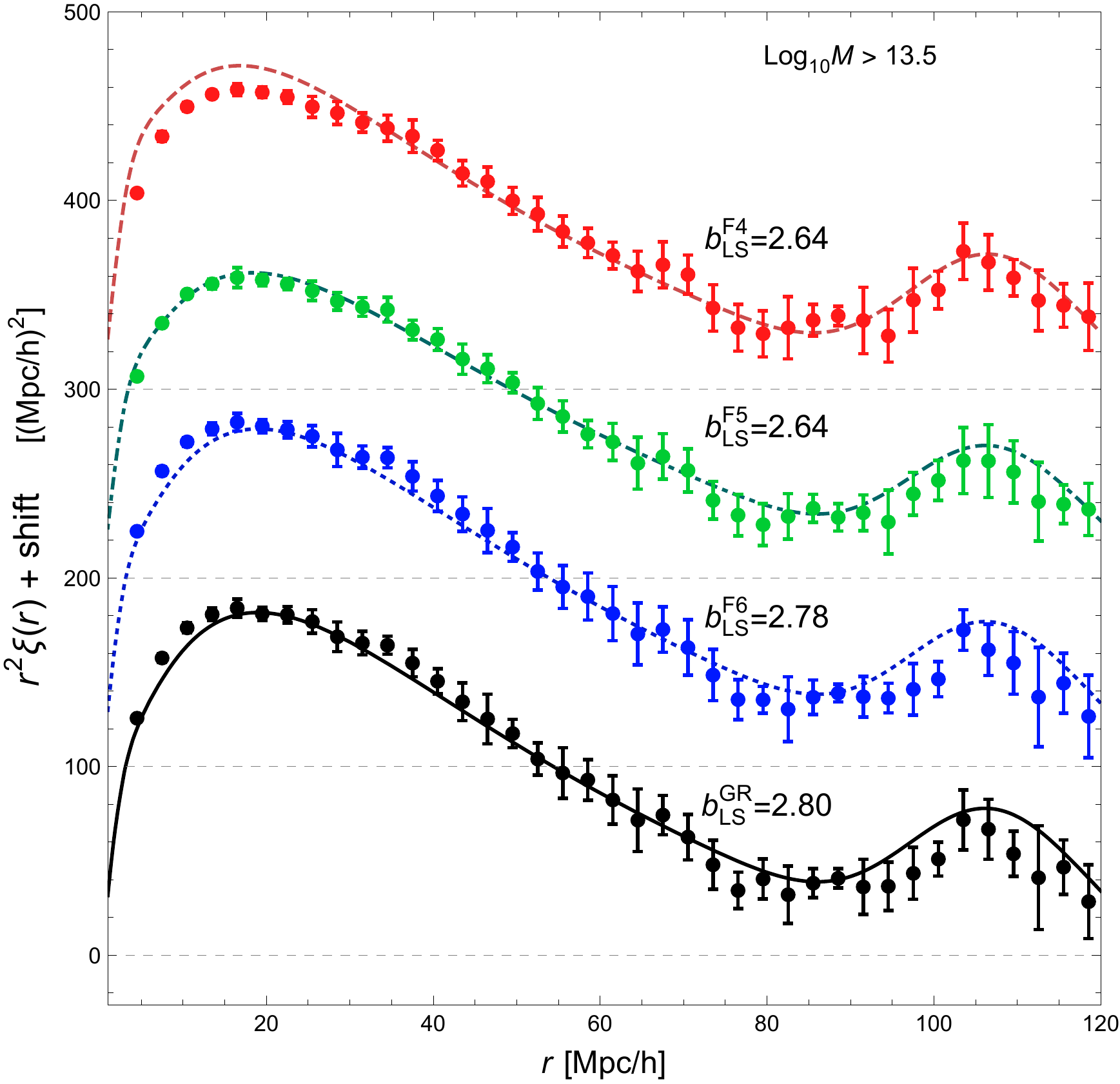}		
	\caption{Halo correlation functions $\xi_h(r)$. From bottom to top, we show GR, F6, F5, and F4 models. Left panel is for halo masses $12.5 < \log_{10} [M/(M_\odot h^{-1})]<15$, and right panel for $13.5 < \log_{10} [M/(M_\odot h^{-1})]<15$. The horizontal dashed gray lines denote the values $\xi_h(r)=0$ for each model, which have been shifted for visualization purposes. The analytical curves are given by $\xi^\text{Model}_h= (1+b_1^\text{Model})^2 \xi^\text{Model}_{m}$, where the matter correlation functions are computed with the ZA. Notice that the Eulerian linear bias (or Large-Scales bias) is $b_\text{LS}=1+b_1$. 
	\label{fig:xish}}
	\end{center}
\end{figure}

\begin{figure}
        \begin{center}
	\includegraphics[width=3 in]{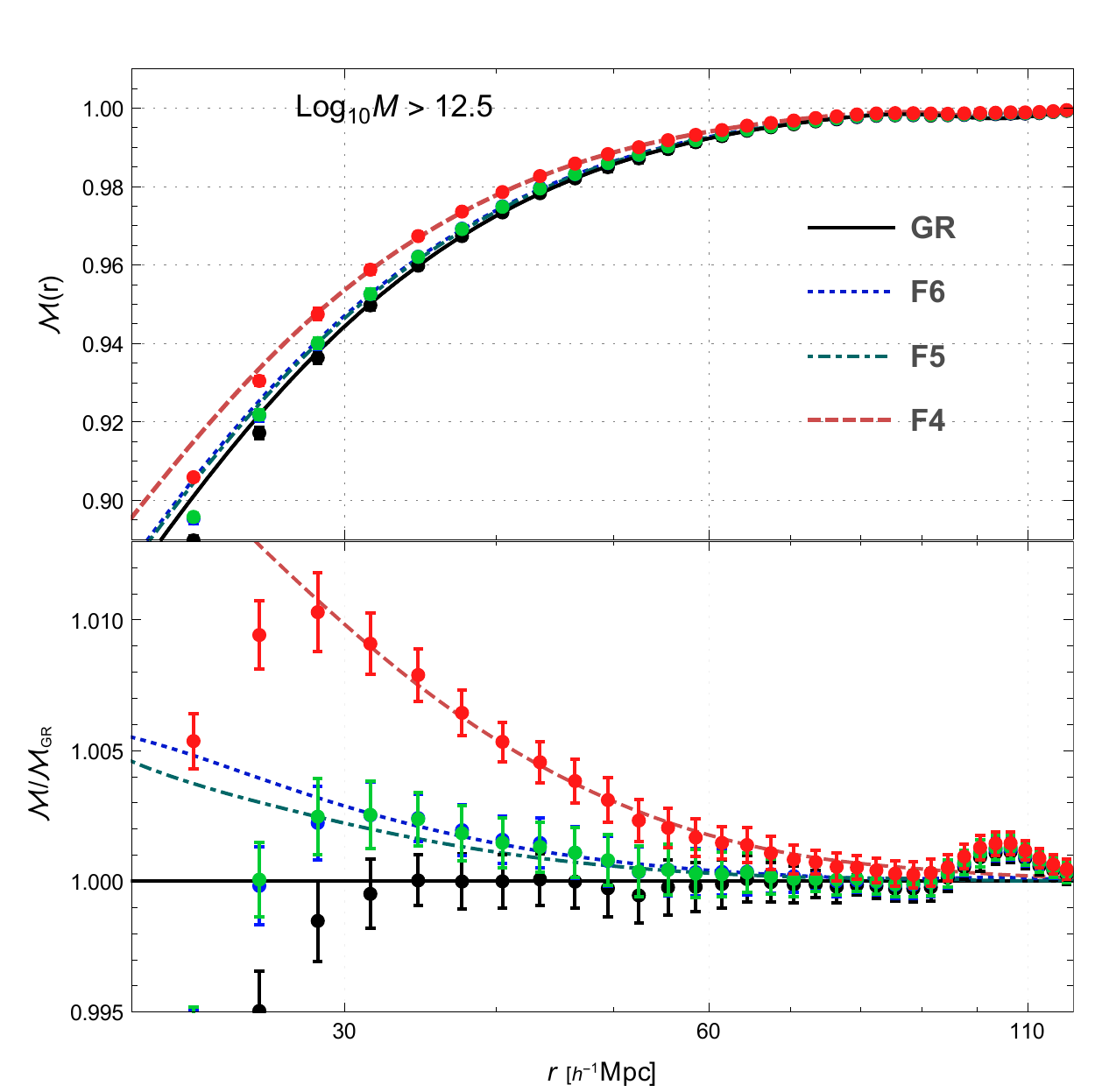}
	\includegraphics[width=3 in]{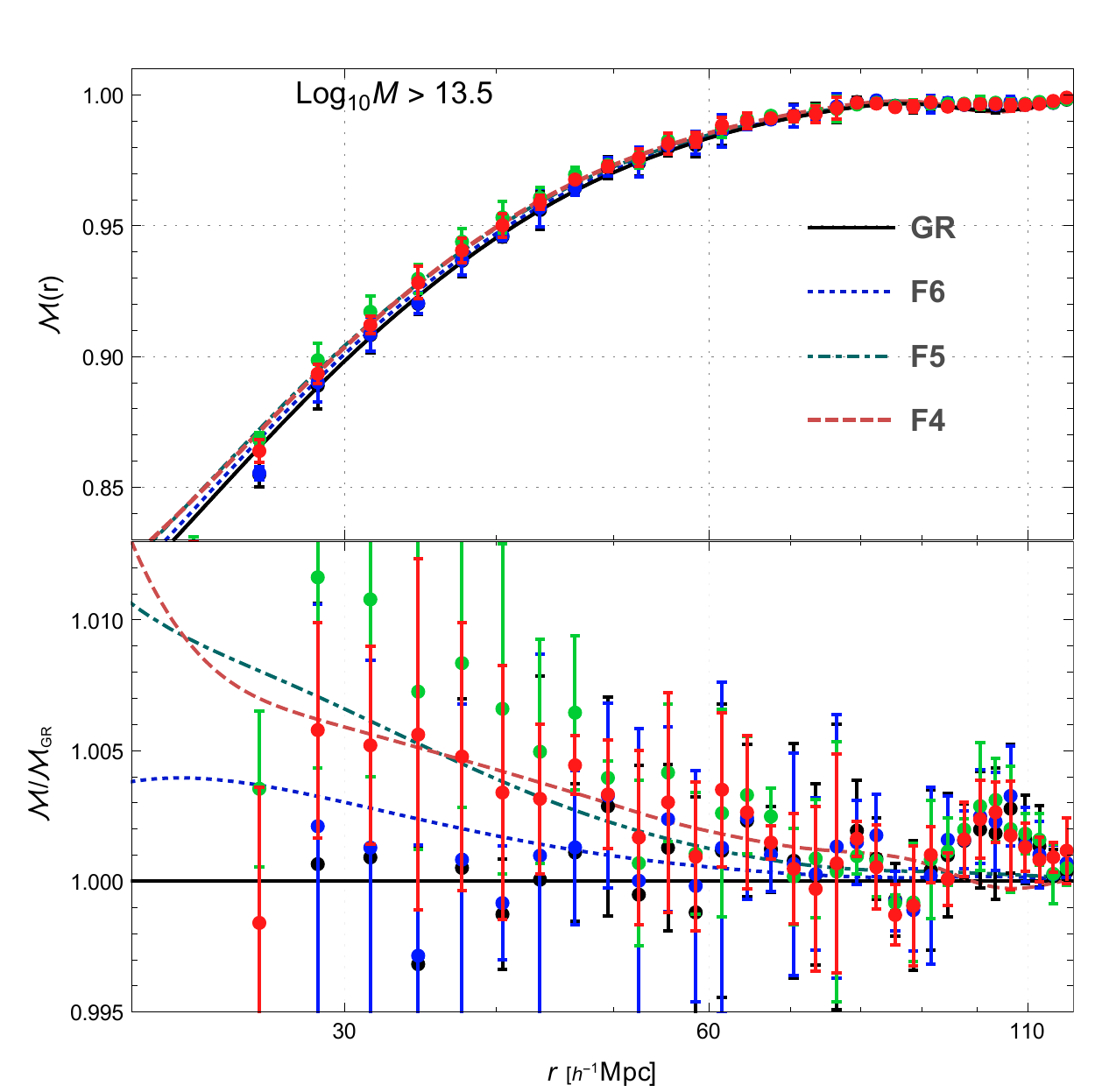}
	\caption{Halo marked correlation functions.  Left panel is for halo masses $12.5 < \log_{10} [M/(M_\odot h^{-1})]<15$, and right panel for $13.5 < \log_{10} [M/(M_\odot h^{-1})]<15$. 
	\label{fig:mCFh} }
	\end{center}
\end{figure}

\begin{figure}
        \begin{center}
	\includegraphics[width=2.8 in]{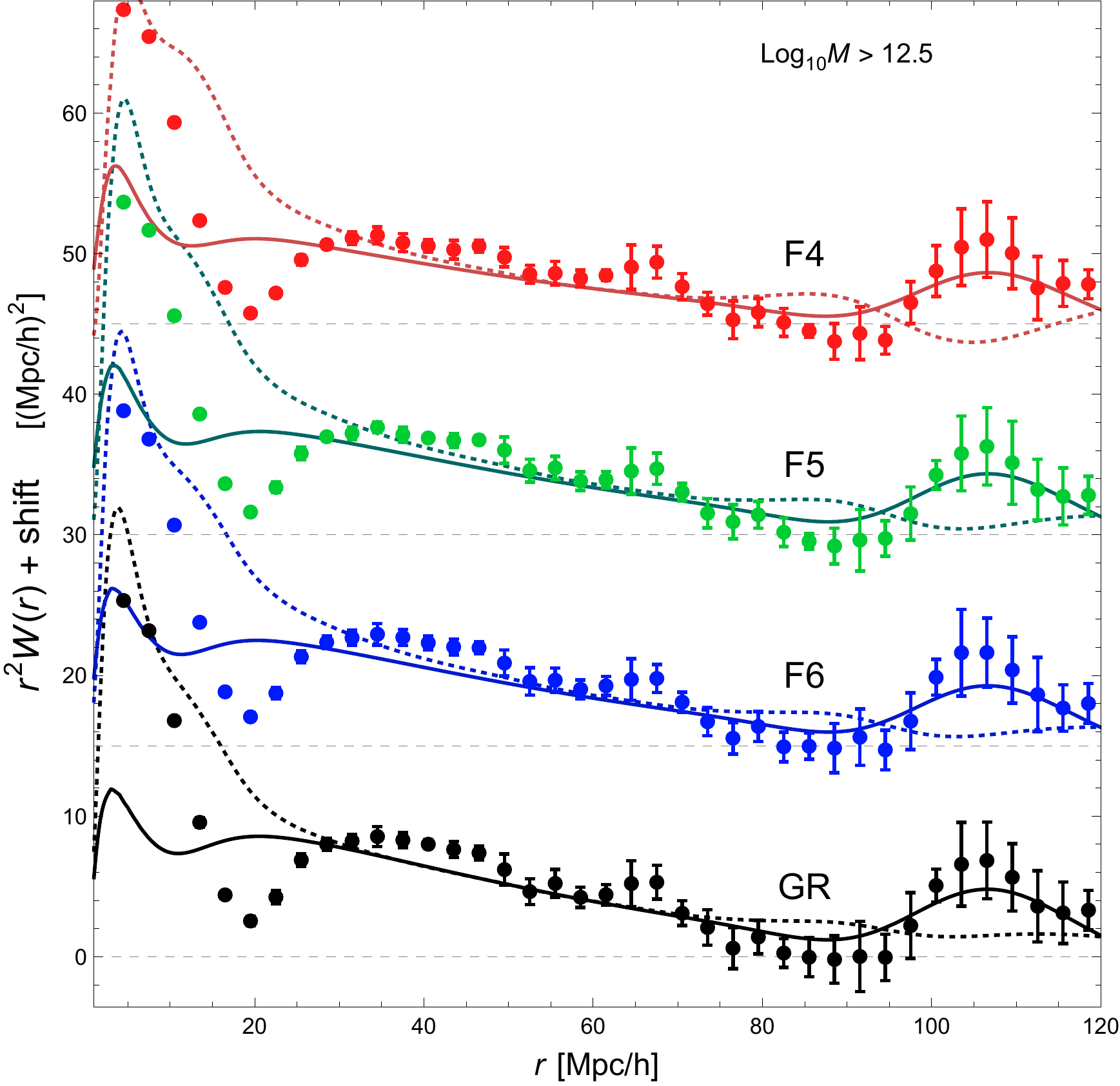}
	\includegraphics[width=2.8 in]{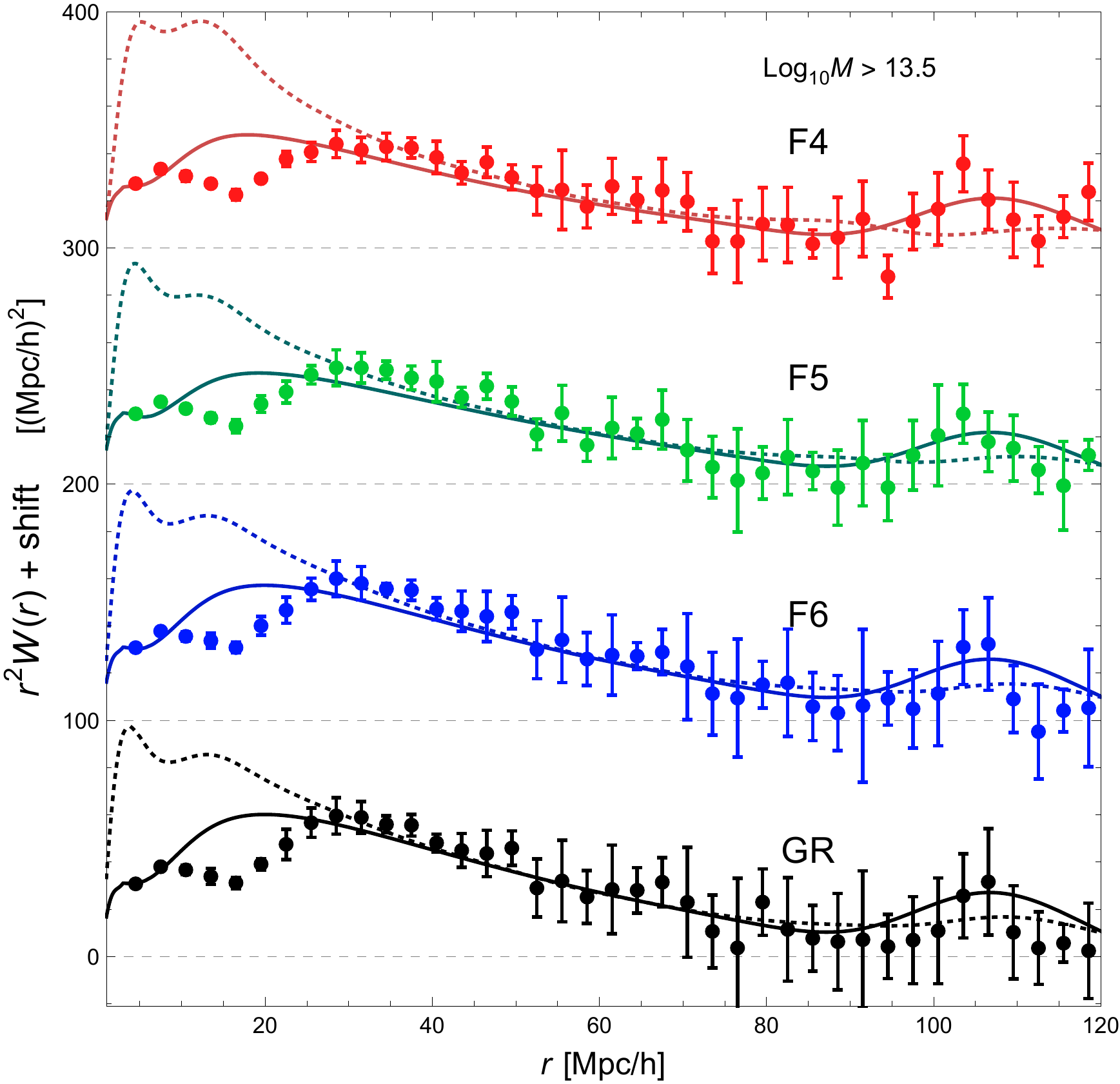}		
	\caption{Halo weighted correlation functions $W_h(r)$.  From bottom to top we show GR, F6, F5, and F4 models. Left panel is for halo masses in the interval $12.5 < \log_{10} [M/(M_\odot h^{-1})]<15$, and right panel for $13.5 < \log_{10} [M/(M_\odot h^{-1})]<15$. The horizontal dashed gray lines denote the zeros for each model, which have been shifted for visualization purposes. The analytical curves are given by the ZA with the bias parameters, $b$, and the expansion parameters, $B$, those obtained from the fitting to the halo correlation functions and to the marked correlation functions, respectively, as it was done in figs.~\ref{fig:xish} and \ref{fig:mCFh}. Dotted curves correspond to the ``This Work'' method and the solid curves to the W16 + 1-loop. 
	\label{fig:Wsh}}
	\end{center}
\end{figure}

Halos are identified from the {\sc elephant} suite of simulations using the publicly available code \texttt{ROCKSTAR} \cite{Behroozi:2011ju}\footnote{https://bitbucket.org/gfcstanford/rockstar} that uses a hierarchical refinement of friends-of-friends groups in phase-space. The particles unbinding procedure used by \texttt{ROCKSTAR} is gravitational model dependent, however in refs.~\cite{Li:2010mqa,Cautun:2017tkc} it is shown that for $f(R)$ theories one can neglect MG effects and use the standard approaches implemented in halo finders.

We test our perturbative method against the halos mCF.
We notice that the mark function is assigned by taking as argument 
the number density of halos, and not the smoothed dark matter density as in eq.~(\ref{MWmark}). That is, we consider $m=m(\delta_R^h)$ with
\begin{equation}
\delta_R^h(\vx) =  
\frac{1}{V_{\Omega_\vx}} \int_{\Omega_\vx}d^3x'  \frac{n_h(\vx') - \bar{n}_h}{\bar{n}_h}, 
\end{equation}
where the region of integration $\Omega_\vx$ corresponds to the the cubic cells of volume $V_{\Omega_\vx}$ used to assign the mark at the position $\vx$. $n_h(\vx)$ is the number density of halos with masses 
in a given interval and $\bar{n}_h$ the average number density of such halos.
The effect of this reassignment is to redefine the mark function, and hence the expansion parameters $B$. In such a case,
we cannot use the values of the $B$ parameters obtained in the previous subsection, and we have to treat them as free parameters that should be fitted from the simulations. To do this, 
we first fit the large-scale, Eulerian bias $b_\text{LS}=1+b_1$ to the correlation function using the ZA, such that 
$\xi_{h}(r) = b_\text{LS}^2 \xi_{\text{ZA}}(r)$. 
We use two interval of masses: $\log_{10} M > 12.5$ and $\log_{10} M > 13.5$, with $M$ the halo mass in units of $M_\odot h^{-1}$. 
In section~\ref{sect:Degeneracies} we discussed that the effect of biasing on the mCF is to bringing it down, such that as more different the biases are, the more the mCF differs. Because of this, we choose a large halo mass interval ($>13.5$) that we expect to give larger different biases for the different gravitational theories.
However, we have a relatively small number of halos in this interval, and hence large statistical errors in the data. Therefore we also use an interval with a moderate lower mass ($>12.5$). A histogram with the number of halos over mass intervals in the {\sc elephant} suite of simulations is presented in figure 2 of ref.~\cite{Hernandez-Aguayo:2018yrp}.

The fittings to the halo correlation functions $\xi_h(r)$  are shown in figure~\ref{fig:xish}, providing the large scale values which are reported in the left and right panels of that figure, corresponding to the two mass intervals considered. As expected, 
the linear local biases are larger for more massive halos, while the error bars also increase because we have 
many fewer halos in the interval  $\log_{10} M > 13.5$ than in the interval $\log_{10} M > 12.5$. Also, we confirm numerically that $b^\text{MG}<b^\text{GR}$, as explained in section \ref{sect:Degeneracies}. However, we do not obtain larger differences in the bias of MG and GR for more massive halos as predicted in the peak-background split formalism. 
This may be an indication of a breakdown of bias models that assume conservation of tracers \cite{Lombriser:2013eza,Aviles:2018saf,Valogiannis:2019xed}: for example, it is known that in F4 fewer small mass halos survive because of a faster merging rate than in GR \cite{Hernandez-Aguayo:2018yrp}. We also note that the number of available halos in the interval  $\log_{10} M > 13.5$ is quite limited due to the limited size of the simulations so the scatter in the correlation function is large.

With the estimation of the biases at hand, we fit the mark function expansion parameters $B$ directly to the mCF, obtaining $B_1^\text{GR}=-1.20$,  $B_1^\text{F6}=B_1^\text{F5}=-1.22$,  $B_1^\text{F4}=-1.25$, for the interval $\log_{10} M > 12.5$; and  $B_1^\text{GR}=-1.43$,  $B_1^\text{F6}=B_1^\text{F5}=-1.45$,  $B_1^\text{F4}=-1.55$, for  $\log_{10} M > 13.5$. We plot these results in figure~\ref{fig:mCFh}, where the top panels show the mCF and the bottom panels the ratio of the mCF of the different gravitational theories to the GR case. 
Our fittings are very good for intermediate scales $30<r<90 \, \text{Mpc}/h$. At larger scales, particularly at the BAO peak, we have a considerable discrepancy. There are two sources of errors, the first one is due to that the used simulations underestimate the BAO peak in the correlation function, see figure~\ref{fig:xish} or refs.~\cite{Valogiannis:2019xed,Valogiannis:2019nfz}; and the second because our analytical model fails at the BAO scale since our method mixes Euleran and Lagrangian correlation functions (see discussion below).  
We have checked that varying $B_2$ and $b_2$ over reasonable intervals has little impact on the results so we let them fixed to $B_2=0.5$ and $b_2=0.2$. Also, the computation of the marked correlation function using the method W16+1-loop gives as accurate results as those shown in figure~\ref{fig:mCFh}, but using slightly different parameters $B$ values. 

To check consistency in our fittings, in figure~\ref{fig:Wsh} we show the weighted correlation function $W_h(r)$ for each model and for each of the halo mass intervals. Dotted curves correspond to the ``This Work'' method and the solid curves to the W16+1-loop. Both methods perform poorly 
at scales $r<30 \,\text{Mpc}/h$, while in the interval $30<r<80 \,\text{Mpc}/h$ the ``This Work'' method works slightly better. However,
at larger scales the method W16+1-loop is superior and it draws correctly the BAO peak. This can be understood analytically by taking the large scale limit for each model,\footnote{See section 3.4 of ref.~\cite{Carlson:2012bu} for the large scale limit of the CLPT correlation function.}
\begin{align}
W^\text{This Work}(r) &\simeq b_\text{LS}^2 \xi_\text{ZA}(r) + B_1^2 \xi_{RR}(r) + 2  b_\text{LS} B_1\xi_{R\sigma_\Psi^2}(r),   \\
W^\text{W16}(r) &\simeq (b_\text{LS} + B_1)^2 \xi_\text{ZA}(r),
\end{align}
where in the second equality we have neglected the differences between $\xi_\text{ZA}$ and its smoothed versions (8th and 12th terms on the right hand side of eq.~(\ref{J0main})) which are smaller than the differences between the ZA and linear Eulerian correlation functions. Hence, W16 is able to capture well the BAO peak because from the very beginning it neglects the differences between Eulerian and Lagrangian coordinates, while our method  fails because the contribution of $\xi_{R\sigma_\Psi^2}$ (being negative for $B_1<0$) competes with the $\xi_{RR}(r)$ and  $\xi_\text{ZA}(r)$ correlation functions.

Finally, we consider the mCF for mock galaxies. 
For the modified gravity models the HOD parameters were fitted as to give rise to the same observed two-point correlation function as in $\Lambda$CDM. 
To test our model to data we proceed exactly as we did for halos above. We first fit the linear local bias, obtaining $b_1^\text{LS}=2.1$ for all models. This is expected of course, because by construction the correlation functions of galaxies should be equal and the matter correlation functions are indistinguishable at large scales. Thereafter, we fit directly to the mCF data. The results are shown in figure~\ref{fig:mCFHOD}, where we only plot the GR and F4 models; the corresponding mCFs for F5 and F6 lie somewhere in between them. 
Given that the correlation function is tuned to be the same in different models, the difference of the mCF is much smaller compared with that for dark matter halos. This implies that the mCF with this choice of the mark is not sufficient to detect the difference between these models. To enhance the difference between the models, we need to consider different marks from White's \cite{Valogiannis:2017yxm,Hernandez-Aguayo:2018yrp,Armijo:2018urs}.

\begin{figure}
        \begin{center}
	\includegraphics[width=3 in]{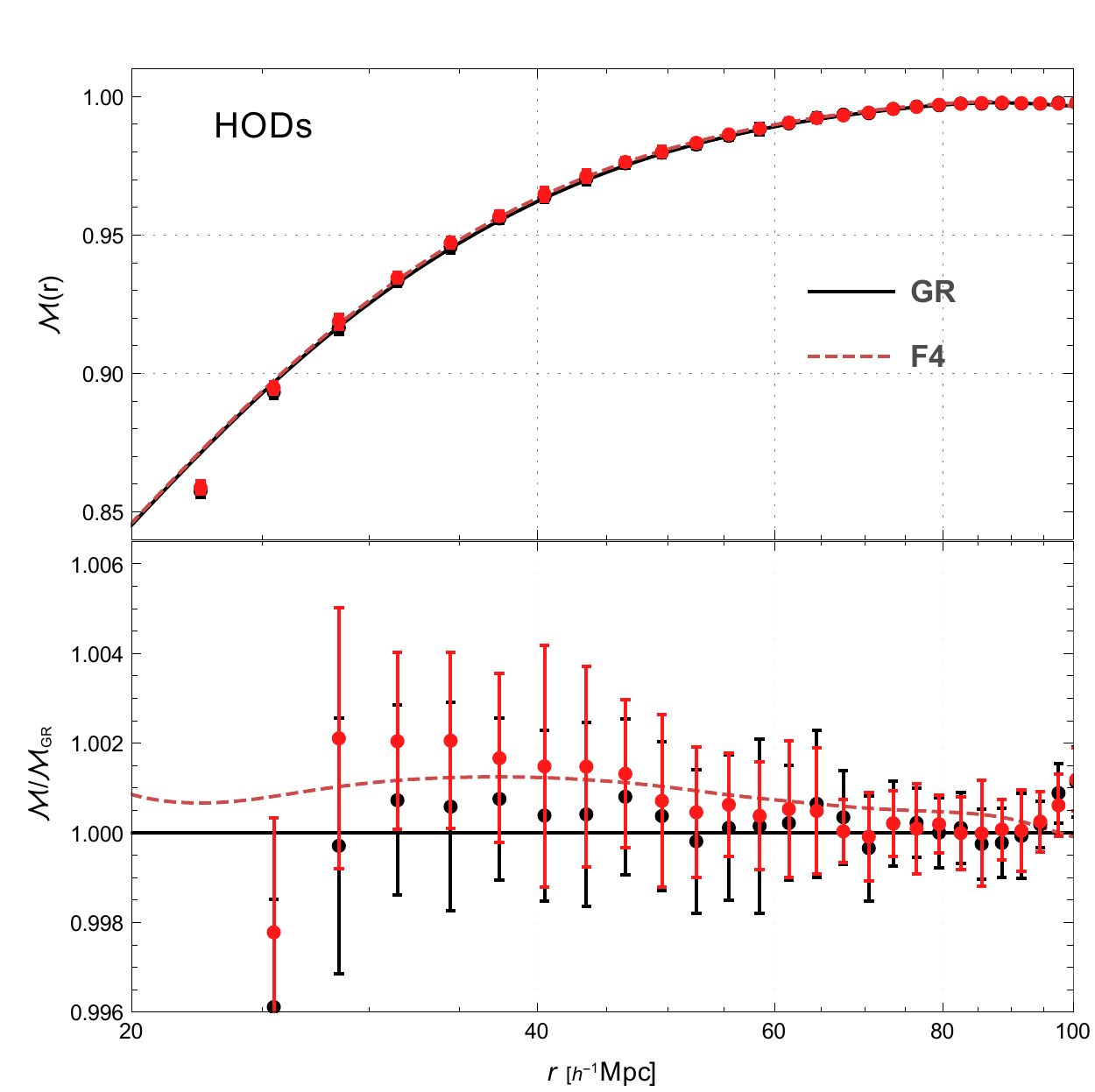}
	\caption{mCFs of HOD galaxy mocks for F4 and GR models. The linear local biases are fitted by using the ZA correlation function and the data from the simulations, giving $b_1^\text{LS}=2.1$ for all models. 
	The $B$ parameters are $B_1^\text{GR}=-1.10$ and $B_1^\text{F4}=-1.05$. 
	\label{fig:mCFHOD}}
	\end{center}
\end{figure}

\end{subsection}

\end{section}

 \begin{section}{Conclusions}\label{sec:Conclu}

 In this work we have studied marked correlation functions that up-weights low density regions in the Universe \cite{White:2016yhs}. The idea behind marked statistics is to assign a value (the mark) to each entity in a point process and perform statistics over the resulting weighted objects. 
The mark can be anything that can be quantified as an astronomical property to be conveniently contrasted such as luminosity, color, morphology, etc..    
A way to suppress nonlinearities in statistics is to choose a mark that gives more weight to objects that reside in low 
density regions, and that smoothly fades away as one moves to higher and higher density regions. This marking technique may be particularly important for testing general relativity with cosmological probes, since modified gravity models, tailored to provide cosmic acceleration, often rely on screening mechanisms to hide their impact on high density environments, and the screening switches off in regions that are more depleted of matter. Therefore, the effects of MG are more pronounced in low density regions. It is then natural to search for such marked statistics, as proposed by Martin White in ref.~\cite{White:2016yhs}.
 
Throughout the paper, we have compared our results with those of ref.~\cite{White:2016yhs}, clarifying some of its issues and generalizing its perturbative model (W16) to include higher than linear order 
fluctuations. We have used renormalized bias parameters and a complete resummation of the mark function's Taylor coefficients in order to write the two-point statistics with no zero-lag correlators. First, we have found an expression for the mCF in linear Eulerian theory, which is the natural frame to assign marks.
However, given that the correlation function is well known to be better described with LPT, we have moved to Lagrangian space, with initially Lagrangian biased tracers, and compute the mCF in the CLPT resummation scheme. This is given by a six dimensional double convolution integral, eq.~(\ref{mCFCLPT2}). We did not 
proceed to directly integrate this complicated integral, but derived two different approximations from it: the one obtained in ref.~\cite{White:2016yhs} plus its loop corrections (W16+loops), and a second that we called throughout as the ``This Work'' model. Both methods have advantages and disadvantages that we 
summarize in the following. In the Lagrangian formulations, one evolves initial, yet linear, fields and assign them the mark at the moment of observation with Eulerian environmental densities. This mix of Eulerian-Lagrangian schemes is difficult to overcome unless one is able to integrate the above-mentioned double convolution. The method W16 from the beginning approximates the Eulerian coordinates to their corresponding Lagrangian coordinates, reducing the expression  of eq.~(\ref{mCFCLPT2}) to a single convolution that can be numerically treated with the known methods of CLPT. Meanwhile, the ``This Work'' method splits the integral into three pieces, one containing only Lagrangian coordinates, one Eulerian, and the third a mix of both. The first two pieces are integrated exactly, but the third, which contains both Lagrangian and Eulerian coordinates, uses the expansion of Lagrangian displacements of eq.~(\ref{approxTW}). When comparing both methods, together with the linear Eulerian and the ZA (the linear contributions to W16; that is, the original model of ref.~\cite{White:2016yhs}), we show that all models give quite similar results, with differences below 1\% for scales $>20 \,\text{Mpc}/h$ and below 0.1\% for $>50 \,\text{Mpc}/h$ (see figure~\ref{fig:CompareModels}); for comparison, the difference between the ZA and 1-loop CLPT correlation functions is about the 1\%  at scales above $r=20\,\text{Mpc}/h$.  This shows the highly linear nature of mCFs that up-weights low density regions of space. At smaller scales, however, the results are unreliable because our perturbative methods are limited by the smoothing scale of the matter (or tracers) densities used to assign the mark, which should be large.
 
We discussed the effects of bias on the mCF, showing that they were quite degenerated with the mark itself. As the mark parameter $B_1$ is more negative, or as larger the value of the linear local bias is, the more the mCF reduces its amplitude. However, as known from earlier works, MG models generally have smaller biases due to a more efficient clustering. Then, while for matter particles the MG mCFs are smaller than that of GR, for biased tracers the opposite may happen. 
 
We tested our analytical methods against the {\sc elephant} suite of MG {\it N}-body simulations. First, we showed results for dark matter. Since this case is free of bias, we can observe the effects of marking dark matter particles more clearly and it allows us to compare the performance of the different perturbative models. 
The four PT methods we tested show different behaviors, but all lie within the 1\% accuracy. However, at smaller scales, $<40\,\text{Mpc}/h$, the method ``This Work'' outperforms the other perturbative approaches and captures pretty well the trend of the data all the way down to the smoothing scale $R=10 \,\text{Mpc}/h$, but it is not accurate enough to lie inside the error bars of the simulation data. 
%
Finally, we compared our PT prediction for biased tracers, dark matter halos and HOD galaxy catalogs, finding good agreement between theory and simulations for scales $r>30 \,\text{Mpc}/h$. However, we notice that at large scales, around the BAO position, the 
W16 method performs better than ``This Work'' 
because of the mixing of Eulerian and ZA correlation functions in our proposed method. 

In this work we formally developed the perturbative theory of marks using dark matter and tracers for general gravity models, and applied to $\Lambda$CDM and modified gravity models to enhance differences among models that eventually could serve to discriminate them with the use of simulations and observations.

 \end{section}

\acknowledgments

We would like to thank Martin White and Georgios Valogiannis for useful discussions and suggestions.
A.A.~and J.L.C.C.~acknowledge support by Conacyt project 283151.
K.K.~is supported by the European Research Council through 646702 (CosTesGrav) and the UK Science and Technologies Facilities Council grants ST/N000668/1 and ST/S000550/1. B.L.~acknowledges supports by the European Research Council via an ERC Starting Grant (ERC-StG-716532-PUNCA) and the UK Science and Technology Facilities Council (STFC) via Consolidated Grant No.~ST/L00075X/1. The simulations described in this work used the DiRAC Data Centric system at Durham University, operated by the Institute for Computational Cosmology on behalf of the STFC DiRAC HPC Facility (\url{www.dirac.ac.uk}). This equipment was funded by BIS National E-infrastructure capital grant ST/K00042X/1, STFC capital grants ST/H008519/1, ST/K00087X/1, STFC DiRAC Operations grant ST/K003267/1 and Durham University. DiRAC is part of the National E-Infrastructure.

 \appendix

\begin{section}{Brief review of LPT for MG}\label{app:LPT}

The LPT for MG was constructed in ref.~\cite{Aviles:2017aor}, in this appendix we give a brief review of that work.
In LPT one considers overdensities $\delta(\vq)$ at an initial time $t_{\rm ini}$ sufficiently early such that $\delta$ is still linear
at all scales of interest of the problem at hand. $\vq$ denote the space coordinates at this initial time and 
the Lagrangian displacement vector $\mathbf{\Psi}$ relates them to Eulerian coordinates defined at later times:
$\vx(\vq,t) = \vq + \mathbf{\Psi}(\vq,t)$, such that  $\vx(\vq,t_{\rm ini}) = \vq$. 
Perturbation theory expands the Lagrangian displacement vector fields  as $\Psi_i(\vk) = \sum_{n=0}^{\infty} \Psi^{(n)}$, and each
term is written in a Fourier space Taylor series in matter linear overdensities as
\begin{equation}
\Psi^{(n)} =\frac{i}{n!}\int 
 \left(\prod_{i=1}^n \Dk{k_i}\right) (2 \pi)^3\dD(\vk-\vk_1-\cdots-\vk_n)
 L_i^{(n)}(\vk_1,...,\vk_n) \delta_L(\vk_1) \cdots \delta_L(\vk_n),
\end{equation}
with $\mathbf{L}^{(n)}$ the Lagrangian kernels at order $n$.
In MG new scales are introduced, such that the linear growth function $D_+(k,t)$ is the fastest growing solution to 
the linearized fluid equation 
\begin{equation}
 (\T-A(k))D_+(k,t) = 0, 
\end{equation}
where $\T \equiv \frac{d^2\,}{dt^2} + 2 H \frac{d\,}{dt}$ \cite{Matsubara:2015ipa}, and we introduced
\begin{align}
 A(k,t) = A_0  \left( 1 + \frac{2 \beta^2 k^2}{k^2 + m^2 a^2} \right), \qquad 
  A_0 = 4 \pi G \bar{\rho}, \label{Af}
\end{align}
with $\bar{\rho}$ the background matter density.
In general $\beta$ and $m$ are time and scale dependent and can be considered as parametrizations 
for unknown MG theories \cite{Bertschinger:2008zb,Silvestri:2013ne}.
Or, otherwise, they can be obtained directly from a specific gravitational model. In scalar tensor theories,  
$\beta$ gives the strength of the fifth-force
and $1/m$ its range. In models with a chameleon screening, the mass $m$ depends on the environmental density becoming large in high density
regions.
Since Lagrangian displacements and matter fluctuations are related by $\delta(\vx) = \big(1-J(\vq)\big)J^{-1}(\vq)$,
with $J_{ij}=\delta_{ij} + \Psi_{i,j}$, at first order we obtain the Zeldovich solution 
\begin{equation}
\Psi_i^{(1)}(\vk,t) = i \frac{k_i}{k^2}D_+(k,t) \delta_L(\vk,t_0), 
\end{equation}
so we can read the first order Lagrangian kernel, $L_i^{(1)}(\vk) = k_i/k^2$. To second order,
\begin{equation}
L_i^{(2)}(\vk_1,\vk_2) = \frac{3}{7} \frac{k_i}{k^2}\left( \mA(\vk_1,\vk_2) - \mB(\vk_1,\vk_2) \frac{(\vk_1\cdot \vk_2)^2}{k_1^2 k_2^2} \right) , 
\end{equation}
with $\vk = \vk_1 + \vk_2$, and
\begin{equation} \label{AandBdef}
 \mA(\vk_1,\vk_2) = \frac{7 D^{(2)}_{\mA}(\vk_1,\vk_2)}{3 D_{+}(k_1)D_{+}(k_2)}, 
 \qquad \mB(\vk_1,\vk_2) = \frac{7 D^{(2)}_{\mB}(\vk_1,\vk_2)}{3 D_{+}(k_1)D_{+}(k_2)},
\end{equation}
and the second order growth functions, $D^{(2)}$, are the solutions to equations \cite{Aviles:2017aor,Winther:2017jof,Aviles:2018saf}
\begin{align}
(\T - A(k))D^{(2)}_{\mA} &= \Bigg[A(k) + (A(k)-A(k_1))\frac{\vk_1\cdot\vk_2}{k_2^2} + (A(k)-A(k_2))\frac{\vk_1\cdot\vk_2}{k_1^2} \nonumber\\
             &     \qquad   -  \left(\frac{2 A_0}{3}\right)^2 \frac{k^2}{a^2} \frac{M_2(\vk_1\vk_2)}{6 \Pi(k)\Pi(k_1)\Pi(k_2)}\Bigg]  D_{+}(k_1)D_{+}(k_2), \label{DAeveq} \\
(\T - A(k))D^{(2)}_{\mB} &= \Big[A(k_1) + A(k_2) - A(k) \Big]  D_{+}(k_1)D_{+}(k_2), \label{DBeveq}
\end{align}
with appropriate initial conditions. As it is common, we have used $\Pi(k) \equiv (k^2 + m^2a^2)/6\beta^2a^2$. The function $M_2$ in 
eq.~(\ref{DAeveq}) is the first coefficient in a Fourier space Taylor expansion of the non-linear piece of the potential in 
the Klein-Gordon equation of the scalar field that mediates the fifth-force \cite{2009PhRvD..79l3512K}, so it is responsible for the screening mechanism that
drives the theory to GR in high density regions. Expressions for the third order growth are large and not displayed here; see ref.~\cite{Aviles:2017aor}.

We stress out that the LPT formalism can be used to obtain statistics in Fourier space as the 2-point correlation function, but also to obtain the power spectrum in SPT, for example, the kernel $F_2$ is given by \cite{Aviles:2018saf}
\begin{align}\label{F2MG}
 F_2(\vk_1,\vk_2) &= \frac{1}{2} + \frac{3}{14}\mA + \left(\frac{1}{2} - \frac{3}{14}\mB \right)\frac{(\vk_1\cdot\vk_2)^2}{\vk_1^2\vk_2^2} 
 + \frac{\vk_1 \cdot \vk_2}{2} \left( \frac{1}{k^2_1} + \frac{1}{k_2^2}\right),
\end{align}
which reduces to the well-known kernel in EdS, since in that case $\mA=\mB=1$.

\subsection{Hu-Sawicki model}\label{app:HS}
The Hu-Sawicki model \cite{Hu:2007nk} is a particular realisation of $f(R)$ gravity that is able to evade the strong constraints coming from local test of gravity and still give rise to interesting observable signatures on cosmological scales. We are here considering the case where the index $n=1$ so the model has only one free parameter $f_{R0}$. Taking this parameter to zero we recover GR. The three choices for the parameters we are considering in this paper (F4, F5 and F6) are such that they lie in the region around where the best constraints lie today. The F4 model corresponds to $|f_{R0}| = 10^{-4}$ (is in tension with local experiments), the F5 model to $|f_{R0}| = 10^{-5}$ (agrees with most experiments and observations, but are in tension with others) and the F6 model to $|f_{R0}| = 10^{-6}$ (which is still allowed).

In this model the functions describing the first order LPT are given by $\beta^2 = 1/6$ and $m(a)=\sqrt{M_1(a)/3}$ where  
\begin{align}\label{M1fR}
M_1(a) = \frac{3}{2}  \frac{H_0^2}{|f_{R0}|} \frac{(\Omega_{m0} a^{-3} + 4 \Omega_\Lambda)^3}{(\Omega_{m0}  + 4 \Omega_\Lambda)^2}.
\end{align}
The function $M_2(a)$ that enters at second order in LPT is given by
\begin{align}\label{M2fR}
M_2(a) = \frac{9}{4}  \frac{H_0^2}{|f_{R0}|^2} \frac{(\Omega_{m0} a^{-3} + 4 \Omega_\Lambda)^5}{(\Omega_{m0}  + 4 \Omega_\Lambda)^4}.
\end{align}
For a complete description of the model and $f(R)$ gravity in general see refs.~\cite{Hu:2007nk,Koyama:2015vza,Clifton:2011jh} and for more information about the LPT equations in this model see ref.~\cite{Aviles:2017aor}. 

\end{section}

\begin{section}{Constructing the marked correlation function}\label{app:mCF}

Starting from eq.~(\ref{mCFCLPT0}) we use the standard methods of CLPT \cite{Carlson:2012bu} and the cumulant expansion theorem to get biases from spectral parameters with the aid of eqs.~(\ref{Bndef}) and (\ref{bndef}). We obtain eq.~(\ref{mCFCLPT1}) with $1+I=1+I^0+I|_\text{zero-lag}$, where
\begin{align} \label{I0}
& 1 + I^0 =  1+ \frac{1}{2}k_1^ik_2^i A^{\rm loop}_{ij} -\frac{i}{6}  \bar{W} + b_1^2 \xi(|\vq_2-\vq_1|)  + 2b_1B_1 \xi_R(|\vx_2-\vq_1|)
 +B_1^2 \xi_{RR}(|\vx_2-\vx_1|) \nonumber\\
 &\quad - ib_1 (k_1^i - k_2^i) U_i(\vq_2-\vq_1)
 - 2iB_1  k_1^i U^R_i(\vx_2-\vq_1) - i b_1b_2 ( k_1^i -  k_2^i)\xi(|\vq_2-\vq_1|)U_i(\vq_2-\vq_1) \nonumber\\
 &\quad 
 -2i b_1^2 B_1 k_1^i \xi(|\vq_2-\vq_1|)U^R_i(\vx_2-\vq_1) -i (b_1^2+b_2) B_1( k_1^i - k_2^i)\xi_R(|\vx_2-\vq_1|)U_i(\vq_2-\vq_1)\nonumber\\
 & \quad
 -2i B_1B_2 k_1^i \xi_R(|\vx_2-\vq_1|)   U^R_i(\vx_2-\vq_1) \nonumber\\
 & \quad - iB_1^2 b_1 (k_1^i -k_2^i)\xi_{RR}(|\vx_2-\vx_1|)  U_i(\vq_2-\vq_1) - 2iB_1B_2 k_1^i \xi_{RR}(|\vx_2-\vx_1|)  U^R_i(\vx_2-\vq_1) \nonumber\\
 & \quad  +\frac{1}{2} \big[ -b_2 (k_1^ik_1^j+k_2^ik_2^j) + b_1^2 (k_1^ik_2^j+k_2^ik_1^j)\big]  U_i(\vq_2-\vq_1)U_j(\vq_2-\vq_1)  \nonumber\\
 & \quad  -2 B_1 b_1(k_1^i -k_2^i) k_1^j U_i(\vq_2-\vq_1)  U^R_j(\vx_2-\vq_1) 
  - (B_1^2 + B_2)  k_1^ik_1^j U^R_i(\vx_2-\vq_1) U^R_j(\vx_2-\vq_1)  \nonumber\\
 &\quad 
  + ib_2 Z^{2000} +iB_2 Z^{0020} + ib_1^2 Z^{1100}  + iB_1^2 Z^{0011} +2ib_1B_1 Z^{1001} 
+2ib_1B_1 Z^{1010}  \nonumber\\
&\quad
- b_1\bar{A}^{1000} -B_1 \bar{A}^{0010},
\end{align}
which is valid to 1-loop and up to third order in bias expansion (counting $b_1$ and $B_1$ as linear and $b_2$ and $B_2$ as second order).
We have defined 
\begin{align}
 Z^{pqrs} &= \langle \delta^p(\vq_1)\delta^q(\vq_2) \delta^r_R(\vx_1)\delta^s_R(\vx_2) (-k_1^i \Psi_i(\vq_1)-k_2^i \Psi_i(\vq_2))  \rangle_c, \\
 \bar{A}^{pqrs} &= \langle \delta^p(\vq_1)\delta^q(\vq_2) \delta^r_R(\vx_1)\delta^s_R(\vx_2) (-k_1^i \Psi_i(\vq_1)-k_2^i \Psi_i(\vq_2)) (-k_1^j \Psi_j(\vq_1)-k_2^j \Psi_j(\vq_2)) \rangle_c, \\
  \bar{W} &= \langle  (-k_1^i \Psi_i(\vq_1)-k_2^i \Psi_i(\vq_2)) (-k_1^j \Psi_j(\vq_1)-k_2^j \Psi_j(\vq_2))(-k_1^k \Psi_k(\vq_1)-k_2^k \Psi_k(\vq_2))  \rangle_c,
\end{align}
and as usual  \cite{Carlson:2012bu}
\begin{align}
U_i(\vq_a-\vq_b) &\equiv \langle \delta(\vq_a) \Psi_i(\vq_b) \rangle = - i\int \frac{d^3 p}{(2 \pi)^3}e^{\vp\cdot (\vq_a-\vq_b)} \frac{p^i}{p^2} P_L(p),  \\
U_i^R(\vx_a-\vq_b) &\equiv \langle \delta_R(\vx_a) \Psi_i(\vq_b) \rangle
= - i\int \frac{d^3 p}{(2 \pi)^3}e^{\vp\cdot (\vx_a-\vq_b)} \frac{p^i}{p^2} P_L(p) \tilde{W}(p). \label{UiR}
\end{align}
In deriving eq.~(\ref{I0}) we have used repeated times that the integral in eq.~(\ref{mCFCLPT0}) is invariant 
under the interchange $(\vk_1,\vx_1,\vq_1)\longleftrightarrow (\vk_2,\vx_2,\vq_2)$. 
Besides $I_0$, there are also terms depending on $\vx_1-\vq_1$ (or $\vx_2-\vq_2$) that lead to zero-lag correlators $\sigma^2_{R\sigma_\Psi}$ after integration, and ultimately
cancel out with the factor $\bar{m}^{-2}$ in the mCF, as it will become more clear in subsection \ref{app:mCFapprox}, 
hence we omit to explicitly write all of them. To linear order in $P_L$, these are
\begin{align}\label{I0zl}
I|_\text{zero-lag} = 2 b_1 B_1 \xi_R(|\vx_1 - \vq_1|) - 2 i B_1 k_1^i U^R_i(\vx_1-\vq_1) + \text{Non-linear}.  
\end{align}

\begin{subsection}{The CLPT marked correlation function}
Using the variables transformation $(\vq_1,\vq_2,\vx_1,\vx_2,\vk_1,\vk_2) \longrightarrow (\vq,\ve Q,\vr,\ve R,\vk_a,\vk_b)$, given in eqs.~(\ref{coordTrans}), into eq.~(\ref{mCFCLPT1}), we analytically perform 
two Gaussian integrations, one for  $d^3k_a$ and the other for $d^3k_b$, to obtain eq.~(\ref{mCFCLPT2}) with
\begin{align}\label{I02}
& 1 + \mathcal{I}^0 =  1 - \frac{1}{2} A^{\rm loop}_{ij}G_{ij}  -\frac{1}{6}\Gamma_{ijk} W_{ijk} + b_1^2 \xi(q)  + 2b_1B_1 \xi_R(z)
 +B_1^2 \xi_{RR}(r) -2 b_1 g_i U_i(\vq) \nonumber\\
 &\quad 
  - 2 B_1  g_i U^R_i(\ve z)  +B_1 g_i^C U^R_i(\ve z) -2  b_1b_2  \xi(q)U_i(\vq)g_i
 + b_1^2 B_1  \xi(q)U^R_i(\ve z)g_i^C - 2 b_1^2 B_1  \xi(q)U^R_i(\ve z)g_i \nonumber\\
 & \quad
 -2 (b_1^2+b_2) B_1\xi_R(z)U_i(\vq)g_i 
 -2 B_1B_2  \xi_R(z)   U^R_i(\ve z)g_i + B_1B_2  \xi_R(z)   U^R_i(\ve z)g^C_i\nonumber\\
 & \quad -2B_1^2 b_1  \xi_{RR}(r)  U_i(\vq)g_i - 2B_1B_2  \xi_{RR}(r)  U^R_i(\ve z)g_i  + B_1B_2  \xi_{RR}(|\vx_2-\vx_1|)  U^R_i(\ve z)g^C_i \nonumber\\
 & \quad  -(b_1^2 + b_2)  U_i(\vq)U_j(\vq)G_{ij}  - 4 B_1 b_1 U_i(\vq)  U^R_j(\ve z)G_{ij} - \frac{1}{2}B_1 b_1 U_i(\vq)  U^R_j(\ve z)g^C_i g_i  \nonumber\\
 & \quad   -  (B_1^2 + B_2)   U^R_i(\ve z) U^R_j(\ve z)G_{ij}  
 -  (B_1^2 + B_2)  (-g_i^C g_j^C + \frac{1}{4}G_{ij}^C) U^R_i(\ve z) U^R_j(\ve z) \nonumber\\
&\quad + b_2 g_i U^{\ominus,2000}_i + B_2 g_i U^{\ominus,0020}_i+  b_1^2 g_i U^{\ominus,1100}_i 
+ B_1^2 g_i U^{\ominus,0011}_i + 2 b_1B_1 g_i U^{\ominus,1001}_i  \nonumber\\
&\quad  + 2  b_1B_1 g_i U^{\ominus,1010}_i - \frac{1}{2} B_2 g_i^C U^{\oplus,0020}_i -\frac{1}{2} B_1^2 g_i^C U^{\oplus,0011}_i 
- b_1B_1 g_i^C U^{\oplus,1001}_i - b_1B_1 g_i^C U^{\oplus,1010}_i \nonumber\\
&\quad
+ b_1 G_{ij} A^{1000}_{ij} + B_1 G_{ij} A^{\ominus\ominus,0010}_{ij}
-  \frac{1}{4}B_1 G^C_{ij} A^{\oplus\oplus,0010}_{ij} -  B_1 g_i g_j^C A^{\oplus\ominus,0010}_{ij} 
\end{align}
where $\ve z =  \vx_2-\vq_1 = \ve R- \ve Q + \frac{1}{2}\vr + \frac{1}{2} \vq$ and
\begin{align}
g^C_i=  C_{ij}^{-1}(Q_j - R_j), \quad G_{ij}^C=C_{ij}^{-1} -g^C_ig^C_j, \quad \Gamma^C_{ijk} =  C_{\{ij}^{-1}g^C_{k\}} - g^C_ig^C_jg^C_k,
\end{align}
where indices in between brackets $\{\cdots\}$ are cyclically summed, and
\begin{equation}
 \Delta_i^\ominus \equiv \Psi_i(\vq_2) - \Psi_i(\vq_1)=\Delta_i , \qquad \Delta_i^\oplus \equiv \Psi_i(\vq_2) + \Psi_i(\vq_1), 
\end{equation}
\begin{align}
U^{\oplus,pqrs}_i &= \langle \delta^p(\vq_1)\delta^q(\vq_2) \delta^r_R(\vx_1)\delta^s_R(\vx_2) \Delta_i^\oplus \rangle_c \\
U^{\ominus,pqrs}_i &= \langle \delta^p(\vq_1)\delta^q(\vq_2) \delta^r_R(\vx_1)\delta^s_R(\vx_2) \Delta_i^\ominus \rangle_c = U^{pqrs}_{i}\\
  A^{\oplus\oplus,pqrs}_{ij} &=\langle \delta^p(\vq_1)\delta^q(\vq_2) \delta^r_R(\vx_1)\delta^s_R(\vx_2) \Delta^{\oplus}_i\Delta^{\oplus}_j \rangle_c \\
  A^{\oplus\ominus,pqrs}_{ij} &=\langle \delta^p(\vq_1)\delta^q(\vq_2) \delta^r_R(\vx_1)\delta^s_R(\vx_2) \Delta^{\oplus}_i\Delta^{\ominus}_j \rangle_c \\
  A^{\ominus\ominus,pqrs}_{ij} &=\langle \delta^p(\vq_1)\delta^q(\vq_2) \delta^r_R(\vx_1)\delta^s_R(\vx_2)  \Delta^{\ominus}_i\Delta^{\ominus}_j \rangle_c = A^{pqrs}_{ij} \\
  W_{ijk} &=  \langle \Delta^{\ominus}_i\Delta^{\ominus}_j \Delta^{\ominus}_k\rangle_c  
\end{align}
Note that if the functions $\xi$, $U$, $A$, $W$, in eq.~(\ref{I02}) depend only on $\vq$ and $\vr$,
we obtain the standard CLPT approach with only one convolution integral, this is because
the following integrals hold
\begin{align}
\int \frac{d^3Q e^{-\frac{1}{2}(\ve R - \ve Q)^T \mathbf{C}^{-1}(\ve R - \ve Q) }  }{(2\pi)^{3/2}|\mathbf{C}|^{1/2}} = 1, &\qquad
\int \frac{d^3Q e^{-\frac{1}{2}(\ve R - \ve Q)^T \mathbf{C}^{-1}(\ve R - \ve Q) }  }{(2\pi)^{3/2}|\mathbf{C}|^{1/2}} g^C_i = 0, \\
\int \frac{d^3Q e^{-\frac{1}{2}(\ve R - \ve Q)^T \mathbf{C}^{-1}(\ve R - \ve Q) }  }{(2\pi)^{3/2}|\mathbf{C}|^{1/2}} G^C_{ij} = 0, &\qquad
\int \frac{d^3Q e^{-\frac{1}{2}(\ve R - \ve Q)^T \mathbf{C}^{-1}(\ve R - \ve Q) }  }{(2\pi)^{3/2}|\mathbf{C}|^{1/2}} \Gamma^C_{ijk},=0,
\end{align}
which are a consequence that the matrix $C_{ij}$ is a function of $\vq$ only. Indeed we have already used the above integrals to derive eq.~(\ref{I02}), where we 
omitted to write terms, like for example $\Gamma^C_{ijk}\langle \Delta^{\oplus}_i\Delta^{\oplus}_j \Delta^{\oplus}_k\rangle_c$, that vanish when integrated over $Q$. 
Moreover, if we evaluate eq.~(\ref{I02}) in $R=Q$,  we have $g^C_i=G^C_{ij}=\Gamma^C_{ijk}=0$, this is what we do in 
section \ref{subsec:mCFCLPT}, where we approximate 
\begin{equation}
\frac{  e^{-\frac{1}{2}(\ve R - \ve Q)^T \mathbf{C}^{-1}(\ve R - \ve Q) }  }{(2\pi)^{3/2}|\mathbf{C}|^{1/2}} \approx \dD(\ve R-\ve Q),
\end{equation}
to obtain eq.~(\ref{preM}) with
\begin{align}\label{J0}
&1 + \mathcal{J}^0 =  1 - \frac{1}{2} A^{\rm loop}_{ij}G_{ij}  -\frac{1}{6}\Gamma_{ijk} W_{ijk} 
+ b_1^2 \xi(q)  + 2b_1B_1 \xi_R(y) +B_1^2 \xi_{RR}(r) \nonumber\\
&\quad -2 b_1 g_i U_i(\vq)
  - 2 b_2  g_i U^R_i(\ve y) -2  b_1b_2  \xi(q)U_i(\vq)g_i
  - 2 b_1^2 B_1  \xi(q)U^R_i(\ve y)g_i \nonumber\\
 &\quad  
 -2 (b_1^2+b_2) B_1\xi_R(y)U_i(\vq)g_i
 -2 B_1B_2  \xi_R(y)   U^R_i(\ve y)g_i \nonumber\\
 & \quad -2B_1^2 b_1  \xi_{RR}(r)  U_i(\vq)g_i - 2B_1B_2  \xi_{RR}(r)  U^R_i(\ve y)g_i   
 -(b_1^2 + b_2)  U_i(\vq)U_j(\vq)G_{ij} 
 \nonumber\\
 & \quad  - 4 B_1 b_1 U_i(\vq)  U^R_j(\ve y)G_{ij}   -  (B_1^2 + B_2)   U^R_i(\ve y) U^R_j(\ve y)G_{ij} \nonumber\\ 
 &\quad + b_1 G_{ij} A^{\ominus,1000}_{ij} + B_1 G_{ij} A^{\ominus,0010}_{ij}
 + b_2 g_i U^{\ominus,2000}_i + B_2 g_i U^{\ominus,0020}_i+  b_1^2 g_i U^{\ominus,1100}_i 
 \nonumber\\
&\quad 
+ B_1^2 g_i U^{\ominus,0011}_i + 2 b_1B_1 g_i U^{\ominus,1001}_i + 2  b_1B_1 g_i U^{\ominus,1010}_i
\end{align}
with 
 $\ve y =  \frac{1}{2}\vr + \frac{1}{2} \vq$.  If we further substitute $\ve y \rightarrow  \vq$, in the arguments of 
 functions $\xi(\ve y)$, $U(\ve y)$, $A(\ve y)$ we obtain eq.~(\ref{1pWM}).

\end{subsection}

\begin{subsection}{Approximating the CLPT marked correlation function}\label{app:mCFapprox}

Keeping for the moment only linear terms in the function $I_{r,q} $ in eq.~(\ref{mCFCLPTForapp1}) and using eqs.~(\ref{I0}) and (\ref{I0zl}) we write at leading order
\begin{align}\label{Irq}
&\langle G_{1}F_{\vx,1}G_{2}F_{\vx,2} \rangle\big|_{r,q} \ni (B_0^{*})^2 \int \Dk{k_1} \Dk{k_2} d^3q_1 d^3q_2 e^{i\vk_1\cdot(\vx_1-\vq_1)} e^{i\vk_2\cdot(\vx_2-\vq_2)}
 e^{-\frac{1}{2}(\vk_1 +  \vk_2)^2\sigma^2_{\Psi}}  \nonumber\\
& \qquad \times \Big[ 2b_1B_1 \xi_R(|\vx_1-\vq_1|)   - i 2B_1 k_1^i \langle \delta_R(\vx_1)\Psi_i(\vq_1) \rangle  
 + 2 b_1B_1 \xi_R(|\vx_2-\vq_1|)\nonumber\\
&\qquad \quad - i 2 B_1 k_1^i \langle \delta_R(\vx_2)\Psi_i(\vq_1) \rangle + \cdots \,\Big] 
 \Big(1 +  \frac{1}{2}k_1^ik_2^j A^L_{ij}(q) \Big). 
\end{align}
The first two terms within the brackets are integrated to give 
$2(B_0^*)^2(1+b_1)B_1 \sigma^2_{R\sigma_\Psi}+\cdots$. Such terms that have as arguments  $\vx_1-\vq_1$ are canceled out by the 
squared of the mean mark in the mCF. 
Let us consider, as an example, the last term in eq.~(\ref{Irq}), which is the more cumbersome
\begin{align}
I_8(r) &\equiv \int \Dk{k_1} \Dk{k_2} d^3q_1 d^3q_2 e^{i\vk_1\cdot(\vx_1-\vq_1)} e^{i\vk_2\cdot(\vx_2-\vq_2)}
 e^{-\frac{1}{2}(\vk_1 +  \vk_2)^2\sigma^2_{\Psi}}   \nonumber\\
 & \qquad \times \Big[- i  k_1^\ell \langle \delta_R(\vx_2)\Psi_\ell(\vq_1) \rangle \frac{1}{2}k_1^ik_2^j A^L_{ij}(q) \Big] \nonumber\\
&=  -\int \frac{ d^3p_1d^3p_2 }{(2\pi)^{6}} 
e^{-i\vp_1\cdot\vr}e^{i\vp_2\cdot\vr}  e^{-\frac{1}{2}p_1^2\sigma^2_{\Psi}} \tilde{W}(k_1)P_L(k_1)
P_L(p_2) \frac{p_1^\ell (\vp_1-\vp_2)^\ell}{p_1^2} \frac{(\vp_1-\vp_2)^i p_2^i}{p_2^2},
\end{align}
where the second equality is obtained after several manipulations and the use of eqs.~(\ref{AijL}) and (\ref{UiR}).
Using
\begin{equation}
  \frac{\vp_1^\ell (\vp_1-\vp_2)^\ell}{p_1^2} \frac{(\vp_1-\vp_2)^i p_2^i}{p_2^2}
  = \delta_{ij}\frac{p_2}{p_1}\hat{p}_1^i\hat{p}_2^j + \delta_{ij}\frac{p_1}{p_2}\hat{p}_1^i\hat{p}_2^j
  -\delta_{ij}\delta_{mn}\hat{p}_1^i\hat{p}_2^j\hat{p}_1^m\hat{p}_2^m - 1,
\end{equation}
and the solid angle integral identities 
\begin{align}
 \frac{-i}{4\pi} \int d \Omega_{\hat{p}} e^{i \vp \cdot \vr}\hat{p}_i &= j_1(pr) \hat{r}_i, \label{saI1}\\
  \frac{1}{4\pi} \int d\Omega_{\hat{p}} e^{i \vp \cdot \vr}\hat{p}_i \hat{p}_j &= \frac{j_1(pr)}{pr}\delta_{ij} - j_2(pr) \hat{r}_i \hat{r}_j, \label{saI2}
\end{align}
we arrive to
\begin{equation}
 I_8 = \frac{4}{3}\xi_{R\sigma_\Psi}(r)\xi_{L}(r)   +\frac{2}{3}\xi_{R\sigma_\Psi}^{[2,0]}(r)\xi_{L}^{[2,0]}(r) 
 -\xi_{R\sigma_\Psi}^{[1,1]}(r)\xi_L^{[1,-1]}(r)-\xi_{R\sigma_\Psi}^{[1,-1]}(r)\xi_L^{[1,1]}(r),
\end{equation}
which is the nonlinear contribution to $\bar{x}_{U^R}$ in eq.~(\ref{barxUR}). 
Analogous manipulations, now including also nonlinear pieces in $I_{p,q}$, yield all the terms in eq.~(\ref{1pWTW}):
\begin{align}
\bar{\text{x}}_{U_RU_R}(r) &= \int \Dk{k} e^{i\vk \cdot \vr} e^{-\frac{1}{2}k^2\sigma^2_\Psi}  Q_9^{RR}(k),\\
\bar{\text{x}}_{U U_R}(r) &= \frac{1}{2}\int \Dk{k} e^{i\vk \cdot \vr} \left( Q_{9}^{R\sigma^2_\Psi}(k)+Q_{12}^{R\sigma^2_\Psi}(k) \right), \\
\bar{\text{x}}_{\xi U_R}(r) &= \int \Dk{k} e^{i\vk \cdot \vr} Q_{12}^{R\sigma^2_\Psi}(k),\\
\bar{\text{x}}_{\xi_R U}(r) &= \int \Dk{k} e^{i\vk \cdot \vr} \left( Q_{12}^{R\sigma^2_\Psi}(k)-\frac{1}{2} Q_{13}^{R\sigma^2_\Psi}(k) \right),  \\
\bar{\text{x}}_{\xi_R U_R}(r) &= \int \Dk{k} e^{i\vk \cdot \vr} e^{-\frac{1}{2}k^2\sigma^2_\Psi} Q_{12}^{RR}(k),   \\
\bar{\text{x}}_{A^{0010}}(r) &= \frac{6}{7} \int \Dk{k} e^{i\vk \cdot \vr}  
\left(R_{3}^{R\sigma_R}(k)+e^{-\frac{1}{2}k^2\sigma^2_\Psi} \tilde{W}_R(k)R_2(k)+Q_{8}^{R\sigma_R}(k) \right), \\
\bar{\text{x}}_{U^{0020}}(r) &= \frac{3}{7} \int \Dk{k} e^{i\vk \cdot \vr} e^{-\frac{1}{2}k^2\sigma^2_\Psi} Q_{8}^{RR}(k), \\
\bar{\text{x}}_{U^{0011}}(r) &=\frac{3}{7}  \int \Dk{k} e^{i\vk \cdot \vr}  R_3^{RR\sigma_\Psi}(k), \\
\bar{\text{x}}_{U^{1010}}(r) &= \frac{3}{7} \int \Dk{k} e^{i\vk \cdot \vr} Q_{8}^{R\sigma_\Psi}(k),   \\
\bar{\text{x}}_{U^{0110}}(r) &= 
\frac{3}{7} \int \Dk{k} e^{i\vk \cdot \vr}\left( e^{-\frac{1}{2}k^2\sigma^2_\Psi} R_{1+2}^R(k) + R_3^{R\sigma_\Psi}(k)\right), 
\end{align}
with $Q(k)$ and $R(k)$ functions
\begin{equation}\label{Q8RR}
 Q_8^{RR}(k)= \int \Dk{p} \left(\A - \B \frac{(\vp\cdot (\vk - \vp))^2}{p^2|\vk - \vp|^2} \right)\tilde{W}_R(p)P_L(p)\tilde{W}_R(|\vk - \vp|)P_L(|\vk - \vp|),
\end{equation}

\begin{equation}
 Q_8^{R\sigma_\Psi}(k)= \int \Dk{p} \left(\A - \B \frac{(\vp\cdot (\vk - \vp))^2}{p^2|\vk - \vp|^2} \right)e^{-\frac{1}{2}p^2\sigma^2_\Psi}\tilde{W}_R(p)P_L(p) 
 P_L(|\vk - \vp|),
\end{equation}

\begin{align}
 R_3^{R\sigma_\Psi}(k) &=  \int \Dk{p} \left(\A - \B \frac{(\vk\cdot\vp)^2}{p^2k^2} \right)e^{-\frac{1}{2}p^2\sigma^2_\Psi}\tilde{W}_R(p)P_L(p) P_L(k),
\end{align}
\begin{equation}
 R_3^{RR\sigma_\Psi}(k) =  \int \Dk{p} \left(\A - \B \frac{(\vk\cdot\vp)^2}{p^2k^2} \right)
 e^{-\frac{1}{2}|\vk-\vp|^2\sigma^2_\Psi}\tilde{W}_R(p)P_L(p) \tilde{W}_R(k)P_L(k),
\end{equation}

\begin{equation}
 R_{1+2}^R(k) = \int \Dk{p} \left(\A - \B \frac{(\vk\cdot\vp)^2}{k^2p^2} \right) \frac{\vk\cdot(\vk-\vp)}{|\vk-\vp|^2} P_L(k) \tilde{W}_R(p)P_L(p), 
\end{equation}
where in $R$ functions we evaluate $\A,\B(\vk,-\vp)$, while in $Q$ functions $\A,\B(\vp,\vk-\vp)$.

\end{subsection}


\end{section}

\begin{section}{{\it k} and {\it q} functions}\label{app:kqfunctions}

$R(k)$ and $Q(k)$  functions are the building blocks for power spectra for both LPT and SPT. They were introduced in \cite{Matsubara:2007wj} 
for matter statistics, and extended for tracers in ref.~\cite{Matsubara:2008wx}. These are constructed by 
2- and 3-point correlations of Lagrangian displacements, and are well known in the literature; we refer the reader to the above references
for their expressions in $\Lambda$CDM, and to refs.~\cite{Aviles:2017aor,Aviles:2018saf}  for MG. In this section we display those functions
that are not presented in the above references but are necessary for the marked correlation functions.
Defining $r=k/p$, $x=\hat{\vk}\cdot\hat{\vp}$ and $y=1+r^2-2rx$:
\begin{align}
 Q_8^{R}(k)     &=  \frac{k^3}{4\pi^2}\int_0^{\infty} dr \tilde{W}_R(kr)P_L(kr) \int_{-1}^1 dx r^2\left(\A-\B\frac{(r-x)^2}{y}\right)  
                       P_L(k\sqrt{y}), \\
 Q_8^{RR}(k)    &=  \frac{k^3}{4\pi^2}\int_0^{\infty} dr \tilde{W}_R(kr)P_L(kr) \int_{-1}^1 dx r^2\left(\A-\B\frac{(r-x)^2}{y}\right)  
                      \tilde{W}_R(k\sqrt{y})P_L(k\sqrt{y}), \\ 
 Q_5^R(k)         &=  \frac{k^3}{4\pi^2} \int_0^\infty dr \tilde{W}_R(kr) P_L(kr) \int_{-1}^1 d x P_L (k \sqrt{y}) \frac{r^2(1- r x)}{y}
                        \left( \A -  \B\frac{(-r +x)^2}{y}  \right),
\end{align}
with $\A$ and $\B$ functions evaluated as the $Q$ functions of the previous appendix.

Correlations functions in CLPT are constructed from a set of $q$-functions $U(\vq)$, $A(\vq)$ and $W(\vq)$; 
see refs.~\cite{Carlson:2012bu,Vlah:2014nta,Vlah:2015sea} for details. 
The expressions of these $q$-functions in MG are derived in ref.~\cite{Aviles:2018saf}, here we write those $q$-functions not presented in that work.
Splitting the $A(\vq)$ functions in 
irreducible components, $A(\vq)=X(q) \delta_{ij} + Y(q) \hat{q}^i\hat{q}^i$, we have for $A^{0010}(\vq)=A^{0001}(\vq)$ 

\begin{align} 
 X^{0010}(q)=X^{0001}(q)  &= \frac{1}{2\pi^2}\int_0^\infty dk \frac{1}{14}\Bigg[ 2 \tilde{W}_R(k)(R_I-R_2) + 3 \tilde{W}_R(k) R_I j_0(kq) \nonumber\\
                          &\qquad- 3(\tilde{W}_R(k)(R_I + 2 R_2) + 2 R^{R}_{1+2} + 2 Q_5^{R}) \frac{j_1(kq)}{kq} \Bigg],\\
 Y^{0010}(q)=Y^{0001}(q)  &= \frac{1}{2\pi^2}\int_0^\infty dk \left(-\frac{3}{14}\right)(\tilde{W}_R(k)(R_I + 2 R_2) + 2 R^{R}_{1+2} + 2 Q_5^{R}) \nonumber\\
 &\qquad \times  \left( j_0(kq) - 3 \frac{j_1(kq)}{kq}  \right).
\end{align}

\noindent Similarly,  $U$ functions can be written as $U_i(\vq) = \hat{q}_i U(q)$ with
\begin{align}
 U^{0020}(q)=U^{0002}(q)  &= \frac{1}{2 \pi^2}\int_0^{\infty} dk k \left( -\frac{3}{7}\right) Q_8^{RR}(k) j_1(kq), \nonumber\\
 U^{1010}(q)=U^{0101}(q)  &= \frac{1}{2 \pi^2} \int_0^{\infty} dk k \left( -\frac{3}{7}\right) Q_8^{R}(k) j_1(kq),  \nonumber\\
 U^{0011}(q)              &=  \frac{1}{2\pi^2}\int dk k \left(-\frac{6}{7}\right) \tilde{W}_R(k) R_{1+2}^R(k) j_1(kq),  \nonumber\\
 U^{1001}(q)=U^{0110}(q)  &=  \frac{1}{2\pi^2}\int dk k \left(-\frac{3}{7}\right) (R_{1+2}^{R}(k) + \tilde{W}_R(k) R_{1+2}(k)) j_1(kq).
\end{align}
In deriving these equations we made use of the identities (\ref{saI1}) and (\ref{saI2}).

\end{section}

\begin{section}{Adding curvature and tidal bias}\label{subsec:curvbias}

\begin{figure}
        \begin{center}
	\includegraphics[width=3 in]{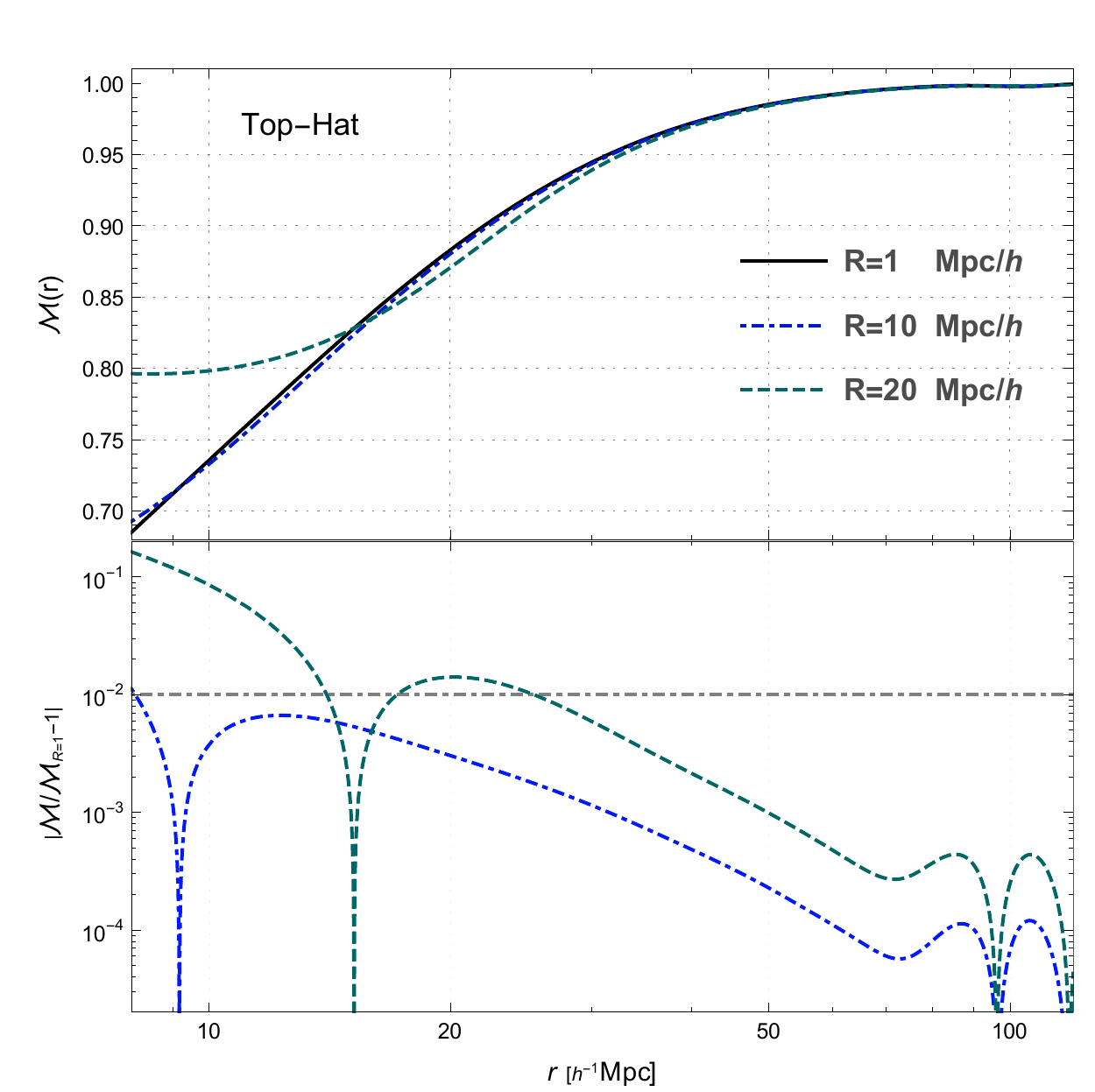}		
	\caption{Marked correlation function for model F5 using a Top-Hat kernel with different smoothing radius $R=1$, $R=10$ and 
	$R=20\, \text{Mpc}/h$. The lower panel shows the relative differences with respect to the $R=1 \,\text{Mpc}/h$ case. The dot-dashed gray horizontal line shows the 1\% offset.
	\label{fig:McompareR}}
	\end{center}
\end{figure}

The non-trivial degeneracy between biases and the mark makes difficult to estimate the mCF
from first principles. On the other hand, the function $W$ has a more neat degeneracy between bias and the mark, indeed  one gets
\begin{equation}
 W(b_1,b_2,B_1,B_2) \approx \xi_X(b_1+B_1, b_2 + 2B_1b_1 + B_2)
 \label{eq:degeneracy}
\end{equation}
where the approximation becomes an equality for $W^\text{W16}$ in the limit $R\rightarrow 0$. 
The identifications $b_1 \rightarrow b_1+B_1$ and $b_2 \rightarrow  b_2 + 2B_1b_1 + B_2$ can be deduced from the definitions of $b_n$ and $B_n$
by replacing $\lambda \rightarrow \lambda + \Lambda$. 
In figure~\ref{fig:McompareR} we use the F5 model to plot the mCF for different smoothing radius $R=1$, $R=10$ and 
$R=20\, \text{Mpc}/h$, using a Top-Hat kernel. We note that for sufficiently large scales
the results are very robust against changes in $R$. 
This observation allows us to use standard methods to introduce curvature bias into the perturbation theory of marks by utilising the relation (\ref{eq:degeneracy}). Perhaps the simplest method to do so is by adding 
\begin{equation}
 2(1+b_1+B_1)b_{\nabla^2\delta} \text{x}_{\nabla^2}(r) + b_{\nabla^2\delta}^2 \text{x}_{\nabla^4}(r), 
\end{equation}
to the $W$ function, and 
\begin{equation}
 2(1+b_1)b_{\nabla^2\delta} \text{x}_{\nabla^2}(r) + b_{\nabla^2\delta}^2 \text{x}_{\nabla^4}(r), 
\end{equation}
to the correlation function of tracers $\xi_X^\text{CLPT}$ \cite{Aviles:2018saf}, with
\begin{align}
\text{x}_{\nabla^2} &=\int \frac{d^3 q}{(2\pi)^{3/2}|\mathbf{A}_L|^{1/2}} e^{-\frac{1}{2}(\vr-\vq)^T\mathbf{A}^{-1}_L(\vr-\vq)}\nabla^2\xi(q), \\
\text{x}_{\nabla^4} &=\int \frac{d^3 q}{(2\pi)^{3/2}|\mathbf{A}_L|^{1/2}} e^{-\frac{1}{2}(\vr-\vq)^T\mathbf{A}^{-1}_L(\vr-\vq)}\nabla^4\xi(q).
\end{align}
Analogously, this prescription can be used to add tidal bias with the standard methods of CLPT (see ref.~\cite{Vlah:2016bcl}  and its extension to MG in ref.~\cite{Valogiannis:2019nfz}): 
whenever the linear bias parameter, $b_1$, appears  accompanying the tidal bias in the correlation function $\xi$, substitute it by  $b_1+B_1$ to obtain the tidal bias contributions to the weighted correlation function $W$.

We have introduced curvature bias for theoretical consistency in MG models. However, we find that for reasonable values of $b_{\nabla^2\delta}$ its influence on the mCF is negligible above a smoothing scale $R=10\,\text{Mpc}/h$, which is the scale we use in section~\ref{sec:CS} to compare the perturbation theory models to simulations. 

 \end{section}



 \bibliographystyle{JHEP}  
 \bibliography{refs.bib}  
 
\end{document}